\documentclass[apjl]{emulateapj}
\usepackage{psfig,amsfonts,amsmath,graphicx,natbib,apjfonts,lscape}

\def\ra#1#2#3{#1$^{\rm h}$#2$^{\rm m}$#3$^{\rm s}$}
\def\dec#1#2#3{$#1^\circ#2'#3''$}
\def\nod{\nodata}

\shorttitle{HST Observations of Short GRB Host Galaxies}
\shortauthors{Fong et al.}

\def\cfa{1}
\def\psu{2}

\begin{document}

\title{{\it Hubble Space Telescope} Observations of Short GRB Host
Galaxies: Morphologies, Offsets, and Local Environments}

\author{ 
W.~Fong\altaffilmark{\cfa}, 
E.~Berger\altaffilmark{\cfa},
\& D.~B.~Fox\altaffilmark{\psu} 
}

\altaffiltext{\cfa}{Harvard-Smithsonian Center for Astrophysics, 60
Garden Street, Cambridge, MA 02138}

\altaffiltext{\psu}{Department of Astronomy and Astrophysics,
Pennsylvania State University, 525 Davey Laboratory, University Park,
PA 16802}

\begin{abstract} We present the first comprehensive analysis of {\it
Hubble Space Telescope} ({\it HST}) observations of short-duration
gamma-ray burst (GRB) host galaxies.  These observations allow us to
characterize the galactic and local environments of short GRBs as a
powerful constraint on the nature of their progenitors.  Using the
{\it HST} data for 10 short GRB hosts we determine the host
morphological properties, measure precise physical and host-normalized
offsets relative to the galaxy centers, and study the locations of
short GRBs relative to their host light distributions.  We find that
most short GRB hosts have exponential disk profiles, characteristic of
late-type galaxies, but with a median size that is twice as large as
that of long GRB hosts, commensurate with their higher luminosities.
The observed distribution of projected physical offsets, supplemented
by ground-based measurements, has a median of $\approx 5$ kpc, about 5
times larger than for long GRBs, and in good agreement with predicted
offset distributions for NS-NS binary mergers.  For the short GRB
population as a whole we find the following robust constraints: (i)
$\gtrsim 25\%$ have projected offsets of $\lesssim 10$ kpc; and (ii)
$\gtrsim 5\%$ have projected offsets of $\gtrsim 20$ kpc.  We find no
clear systematic trends for the offset distribution of short GRBs with
and without extended soft emission.  While the physical offsets are
larger than for long GRBs, the distribution of host-normalized offsets
is nearly identical due to the larger size of short GRB hosts.
Finally, unlike long GRBs, which are concentrated in the brightest
regions of their host galaxies, short GRBs appear to under-represent
the light distribution of their hosts; this is true even in comparison
to core-collapse and Type Ia supernovae.  Based on these results, we
conclude that short GRBs are consistent with a progenitor population
of NS-NS binaries, but partial contribution from prompt or delayed
magnetar formation is also consistent with the data.  Our study
underscores the importance of future {\it HST} observations of the
larger {\it existing} and growing sample of short GRB hosts, which
will allow us to delineate the properties of the progenitor
population.
\end{abstract}


\section{Introduction}
\label{sec:into}

The galactic and local environments of cosmic explosions provide
powerful insight into the nature of their progenitors.  For example,
past studies of supernova (SN) environments have demonstrated that
Type Ia and Type Ib/Ic/II events arise from distinct progenitor
systems since the former are located in all types of galaxies, while
the latter occur only in star forming galaxies, pointing to a direct
link with massive stars (e.g., \citealt{vlf05}).  In a similar vein,
long-duration gamma-ray bursts (GRBs; duration, ${\rm T90}\gtrsim 2$
s) have been linked with massive stars through their exclusive
association with star forming galaxies (e.g.,
\citealt{bdk+98,dkb+98,ftm+99}).  Short-duration GRBs, on the other
hand, are now known to reside in all types of galaxies
\citep{bpc+05,ffp+05,gso+05,bpp+06,ber09}.  Moreover, even the star
forming host galaxies of short GRBs differ from those of long GRBs;
they have higher luminosities and metallicities, and lower specific
star formation rates \citep{ber09}.  The difference between long and
short GRB host galaxies, along with the lack of SN associations for
short GRBs \citep{bpc+05,ffp+05,gso+05,hwf+05,bpp+06,sbk+06},
demonstrate that they have distinct progenitor populations.  In
particular, at least some short GRBs are associated with an older
progenitor population.

Equally important are the local, sub-galactic environments.  In the
case of long GRBs, the distribution of their projected physical and
host-normalized offsets relative to the host centers matches the
overall expected distribution for massive stars in an exponential disk
\citep{bkd02}.  An analysis of the brightness distribution at the
locations of long GRBs indicates that they are disproportionately
concentrated on the brightest regions of their hosts, primarily in
comparison to core-collapse SNe, which follow the overall light
distribution of their hosts \citep{fls+06}.  Both of these studies
have relied on high angular resolution {\it Hubble Space Telescope}
({\it HST}) imaging observations.  Preliminary studies of short GRB
offsets \citep{bpc+05,ffp+05,bp06,sbk+06,tko+08,dmc+09} reveal
somewhat larger projected physical offsets than for long GRBs, and
have also led to a claimed trend of smaller offsets for short GRBs
with extended X-ray emission \citep{tko+08}.  No study of the
locations of short GRBs relative to their hosts light distribution has
been published so far.

Progenitor models of short GRBs lead to distinct predictions about the
distribution of host properties and the local environments of short
GRBs (measured by their offsets and location relative to the host
light distribution).  In particular, the popular model of neutron star
and/or black hole binary mergers (NS-NS/NS-BH; \citealt{elp+89,npp92})
predicts larger offsets than for the massive star progenitors of long
GRBs due to potential systemic velocity kicks.  Various authors have
employed population synthesis models to predict the distribution of
offsets by convolving distributions of kick velocities, merger
timescale, and galaxy masses \citep{bsp99,fwh99,bpb+06}.  For Milky
Way mass galaxies, appropriate for short GRB hosts \citep{ber09}, the
predicted offset distributions have a median of $\sim 5-10$ kpc, with
a broad tail extending to tens of kpc.  On the other hand, progenitor
models that invoke magnetars, either from a young population or
through delayed formation in a WD-WD merger or white dwarf
accretion-induced collapse \citep{lwc+06,mqt08}, are expected to have
a modest offset distribution since these systems do not experience
kicks.  Similarly, any systematic differences in the progenitors of
short GRBs with and without extended X-ray emission may be revealed in
the offset distribution and specific sub-galactic environments.

To confront these models with observations we require high angular
resolution imaging, best provided by {\it HST}.  Such observations
provide detailed information on the host galaxy morphological
properties (e.g., exponential disk versus de Vacoulers profile,
effective radius), as well as the ability to precisely measure offsets
and the distribution of short GRBs relative to their host light.  {\it
HST} observations have served as the backbone for detailed studies of
long GRB environments and host galaxy morphologies
\citep{bkd02,fls+06,wbp07}.  To date, {\it HST} observations of only 3
short GRBs have been published (050709: \citealt{ffp+05}; 060121:
\citealt{ltf+06}; 080503: \citealt{pmg+09}), and in only one case
(050709) was the host morphology addressed.

Here we present the first comprehensive analysis of all short GRB host
galaxies observed with {\it HST} to date\footnotemark\footnotetext{We
do not repeat the analysis for GRB\,080503 since no convincing host
galaxy was identified \citep{pmg+09}.  Future {\it HST} ACS or WFPC3
observations will provide better constraints on an underlying host
than available from the existing WFPC2 observations.}.  Using these
observations we determine the host morphologies and structural
properties (\S\ref{sec:morph}), we calculate precise physical and
host-normalized offsets using accurate astrometry relative to
ground-based afterglow images (\S\ref{sec:offsets}), and we construct
the first distribution of short GRB locations relative to their host
light (\S\ref{sec:light}).  We draw conclusions in the context of
progenitor models in \S\ref{sec:disc}.  Throughout the paper we
compare and contrast the results of our analysis with similar studies
of long GRBs.

We find that: (i) the short GRB hosts have systematically larger
effective radii than long GRB hosts, in good agreement with their
higher luminosities; (ii) the observed short GRB projected physical
offset distribution has a median of about 5 kpc, about a factor of 5
times larger than long GRBs, while for the population as a whole
$\gtrsim 25\%$ have projected offsets of $\lesssim 10$ kpc, and
$\gtrsim 5\%$ have projected offsets of $\gtrsim 20$ kpc; (iii) both
the observed physical offset distribution and the robust constraints
closely match the predicted offset distribution of NS-NS binaries; and
(iv) short GRBs uniformly trace the light distribution of their hosts,
similar to core-collapse SNe, but distinct from long GRBs.

\section{Hubble Space Telescope Data Reduction and Analysis}
\label{sec:data}

\subsection{Sample}

We present {\it HST} optical observations of ten short GRB host
galaxies obtained with the Advanced Camera for Surveys (ACS) and the
Wide-Field Planetary Camera 2 (WFPC2).  The data were obtained as part
of programs 10119, 10624, and 10917 (PI: Fox), as well as 10780 and
11176 (PI: Fruchter).  These programs targeted all short GRBs with
optical and X-ray (XRT) positions from May 2005 to December 2006,
which were visible during {\it HST} 2-gyro operations.  In this time
frame, the only short burst that was not observed was GRB\,060801.
Thus, the sample in this paper is nearly complete in relation to the
short GRBs with optical/X-ray afterglows\footnotemark\footnotetext{We
do not include in this analysis GRBs 051227, 060505, and 060614 all of
which had durations well beyond 2 s, even when isolating the initial
``short'' emission episode.  The {\it HST} data for GRBs 060505 and
060614 were published in \citet{ocg+07} and \citet{gfp+06},
respectively.}.  The {\it HST} observations of GRBs 050709 and 060121
have been published previously by \citet{ffp+05} and \citet{ltf+06},
respectively, but we re-analyze them here in a uniform fashion and
perform a more comprehensive analysis of the host morphology and burst
environment.  Seven of the nine bursts have been localized to
sub-arcsecond precision from optical afterglow detections, and five
have known redshifts (Table~\ref{tab:hst}); we use a constraint of
$z\gtrsim 1.4$ for GRB\,051210 \citep{bfp+07}.

Six of the seven short GRBs with sub-arcsecond localization are
robustly associated with host galaxies\footnotemark\footnotetext{For a
complete discussion of the galaxy associations we refer the reader to
the following papers: GRB\,050709: \citet{ffp+05}; GRB\,050724:
\citet{bpc+05}; GRB\,051221: \citet{sbk+06}; GRB\,060121:
\citet{ltf+06}; GRB\,061006: \citet{bfp+07} and \citet{dmc+09}.} (we
present a host identification for GRB 060313 in this paper; see
Appendix~\ref{sec:app1}). The sole exception is GRB\,061201, for which
we explore two possible host galaxy associations based on the {\it
HST} observations (see Appendix~\ref{sec:app2}); previously only one
galaxy (at $z=0.111$) was considered a potential host
\citep{bfp+07,sdp+07}.  Details of the GRB properties and the {\it
HST} observations are provided in Table~\ref{tab:hst}.

Throughout the paper we use the standard cosmological parameters,
$H_0=71$ km s$^{-1}$ Mpc$^{-1}$, $\Omega_m=0.27$, and
$\Omega_\Lambda=0.73$.  All reported magnitudes are corrected for
Galactic extinction using \citet{sfd98} dust maps.

\subsection{Hubble Space Telescope Data Reduction}

\begin{figure}
\centering
\includegraphics[angle=0,width=5.0in]{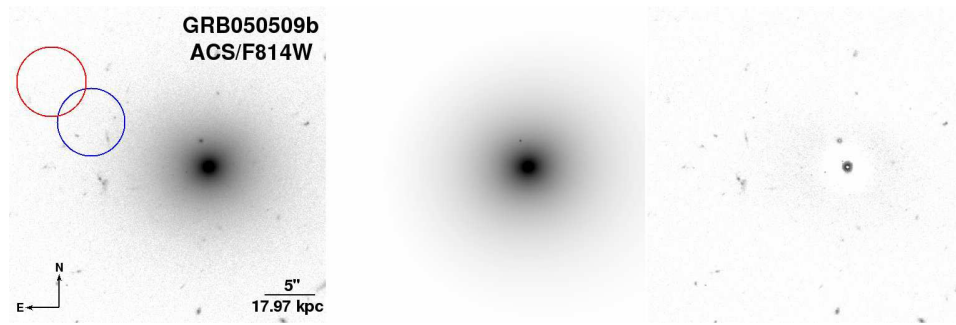}
\caption{{\it Left:} {\it HST}/ACS/F814W image of the location of
GRB\,050509b.  The circles mark the X-ray positions of the afterglow
from the analysis of \citet{but07} (red) and \citet{ebp+09} (blue).
{\it Center:} S\'{e}rsic model fit from {\tt galfit}. {\it Right:}
Residual image.}
\label{fig:050509b}
\end{figure}

\begin{figure}
\centering
\includegraphics[angle=0,width=5.0in]{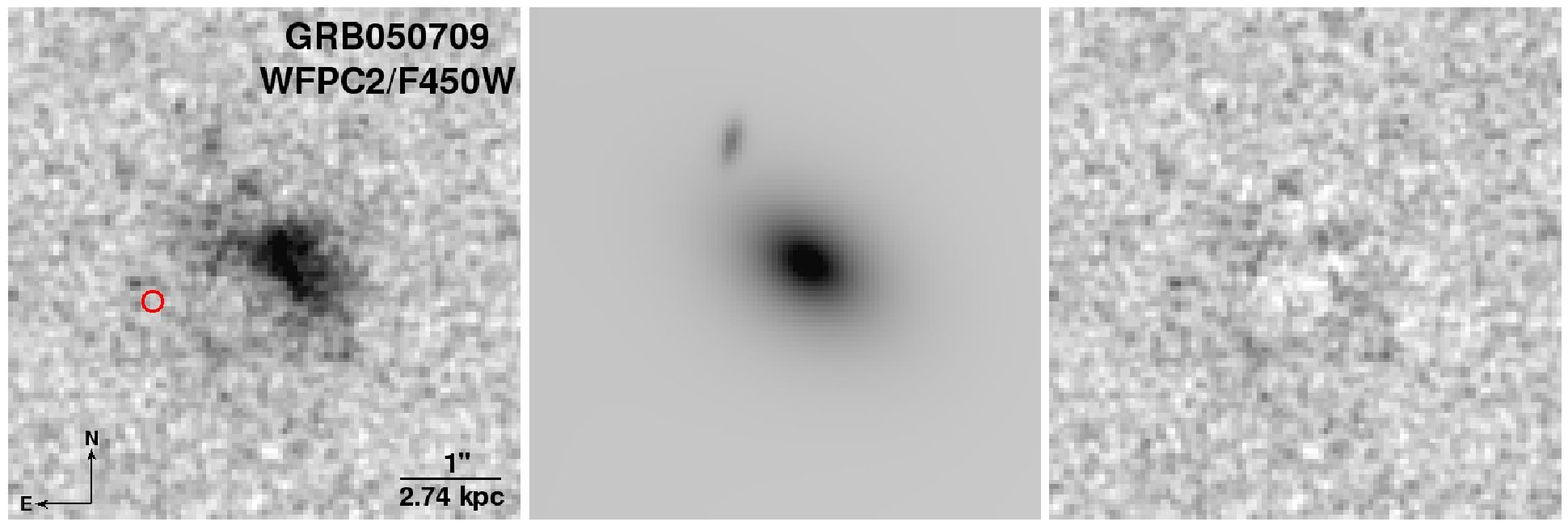}
\includegraphics[angle=0,width=5.0in]{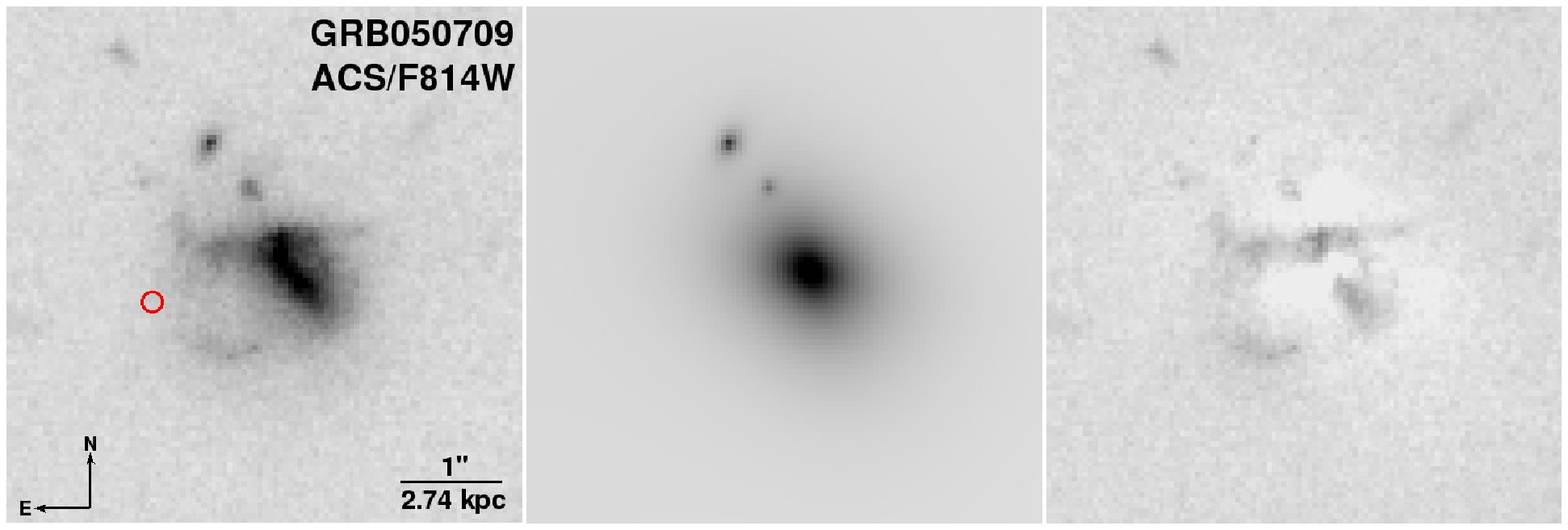}
\caption{{\it Top-left:} {\it HST}/ACS/F450W image of the host galaxy
of GRB\,050709 with a $10\sigma$ error circle representing the
afterglow position.  {\it Top-center:} S\'{e}rsic model fit from {\tt
galfit}.  {\it Top-right:} Residual image.  {\it Bottom:} Same, but
for the {\it HST}/WFPC2/F814W observations.
\label{fig:050709}}
\end{figure}

\begin{figure}
\centering
\includegraphics[angle=0,width=5.0in]{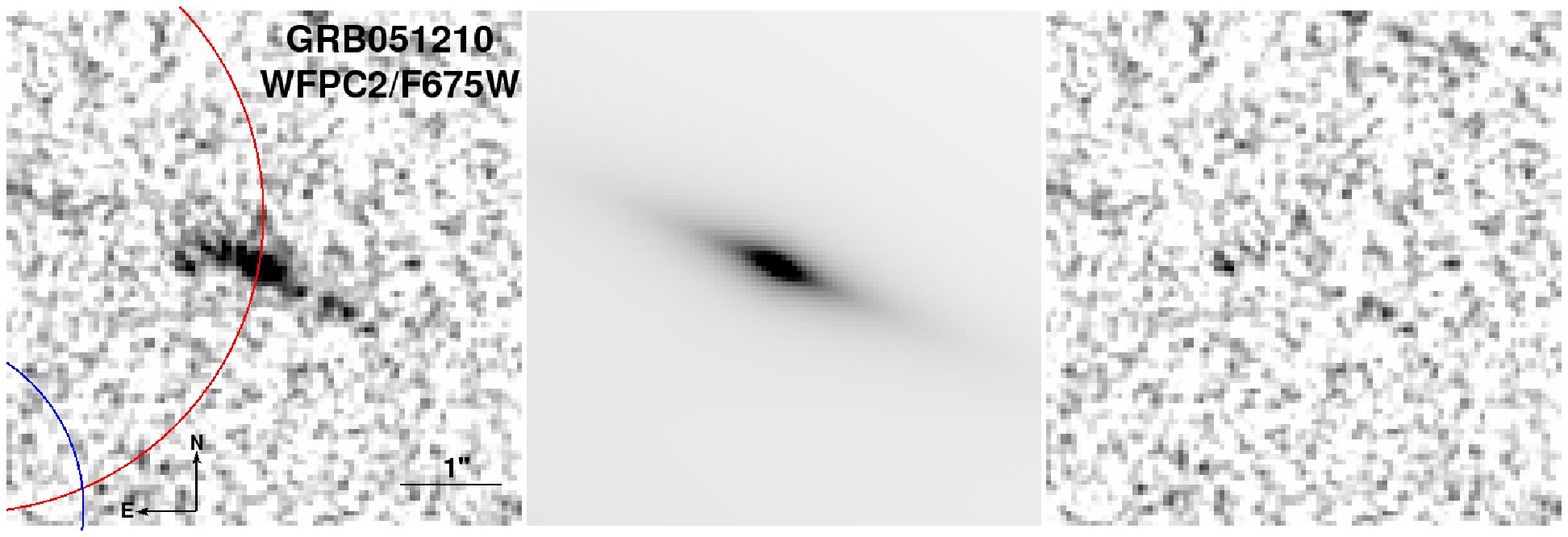}
\includegraphics[angle=0,width=5.0in]{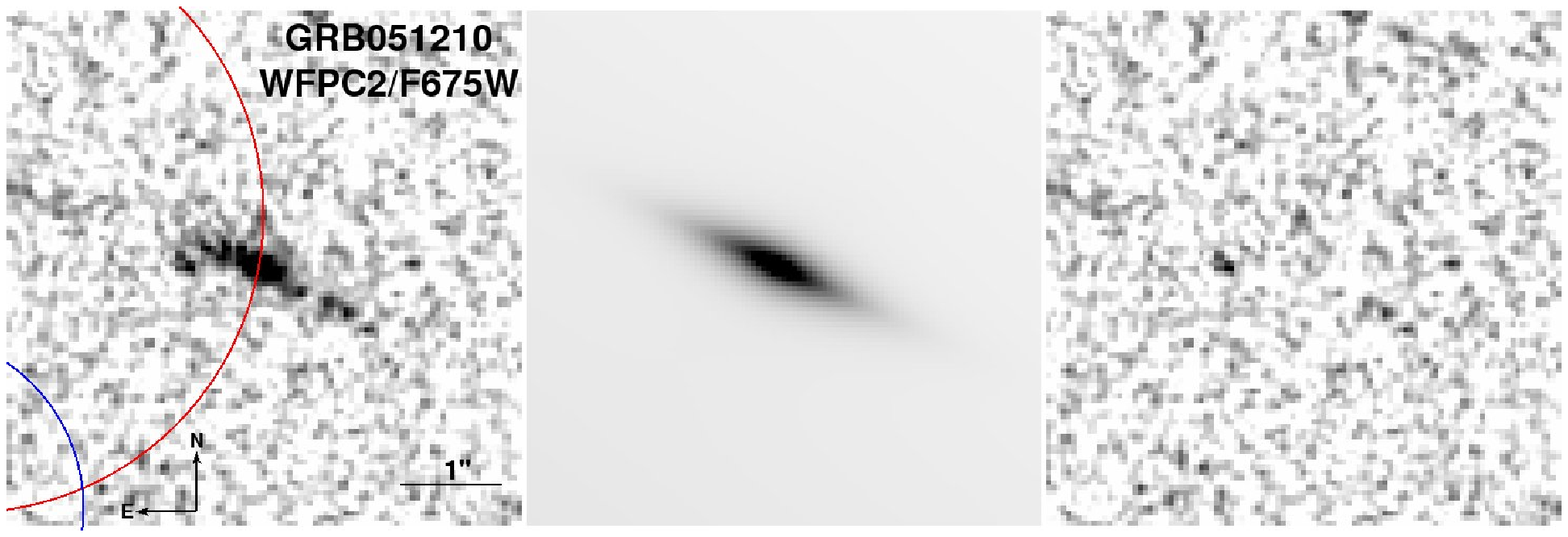}
\caption{{\it Top-left:} {\it HST}/WFPC2/F675W image of the location
of GRB\,051210.  The circles mark the X-ray positions of the afterglow
from the analysis of \citet{but07} (red) and \citet{ebp+09} (blue).
{\it Top-center:} S\'{e}rsic model fit from {\tt galfit} with a fixed
value of $n=1$. {\it Top-right:} Residual image.  {\it Bottom:} Same,
but for S\'{e}rsic model fit from {\tt galfit} with a fixed value of
$n=4$.
\label{fig:051210}}
\end{figure}

\begin{figure}
\centering
\includegraphics[angle=0,width=5.0in]{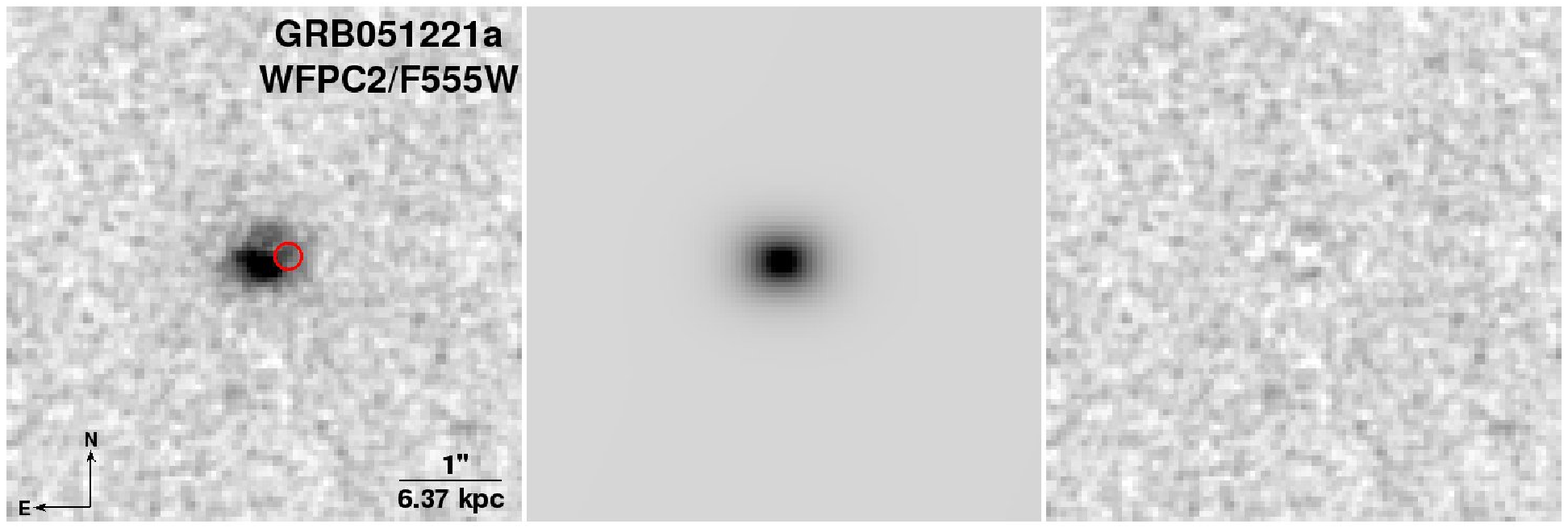}
\includegraphics[angle=0,width=5.0in]{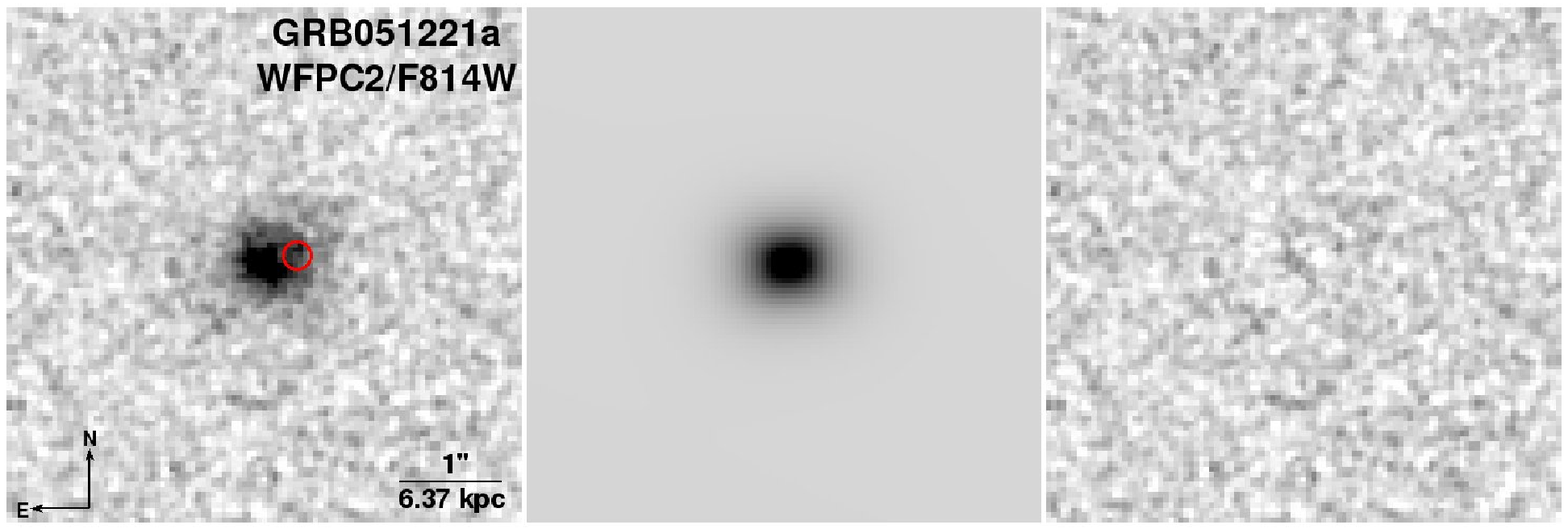}
\caption{{\it Top-left:} {\it HST}/WFPC2/F555W image of the host
galaxy of GRB\,051221 with a $5\sigma$ error circle representing the
afterglow position.  {\it Top-center:} S\'{e}rsic model fit from {\tt
galfit}.  {\it Top-right:} Residual image.  {\it Bottom:} Same, but
for the {\it HST}/WFPC2/F814W observations.
\label{fig:051221}}
\end{figure}

\begin{figure}
\centering
\includegraphics[angle=0,width=5.0in]{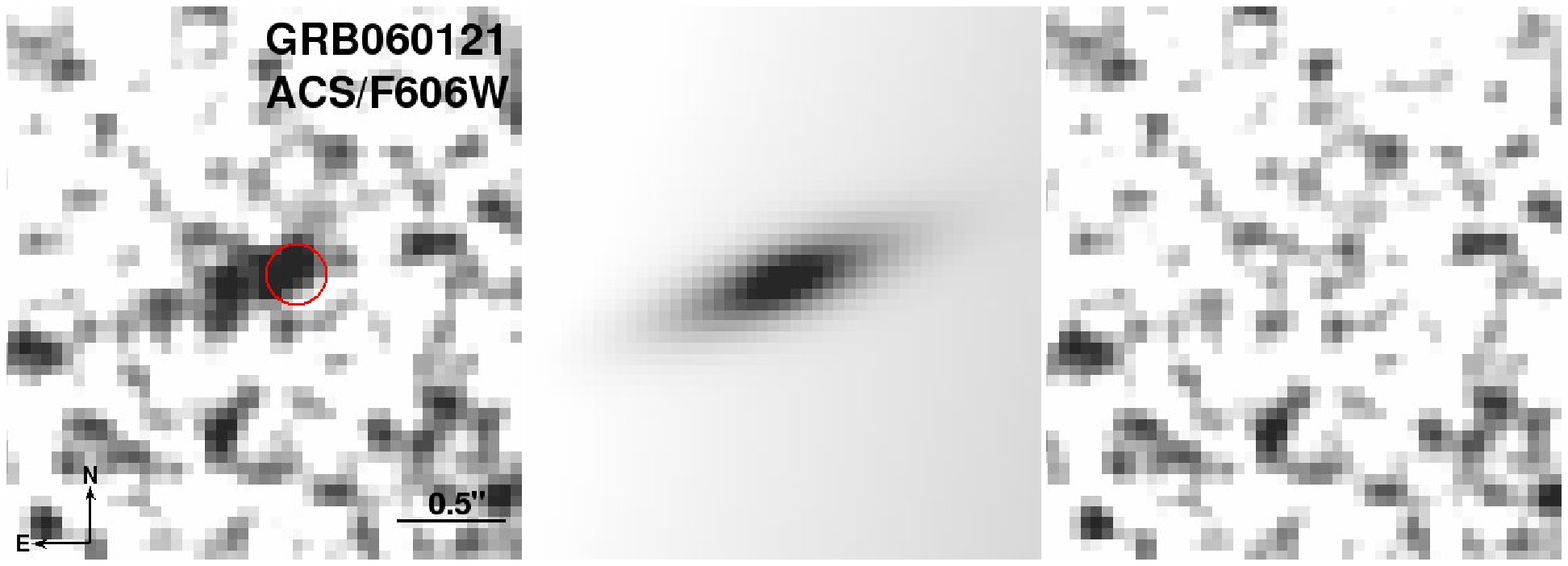}
\includegraphics[angle=0,width=5.0in]{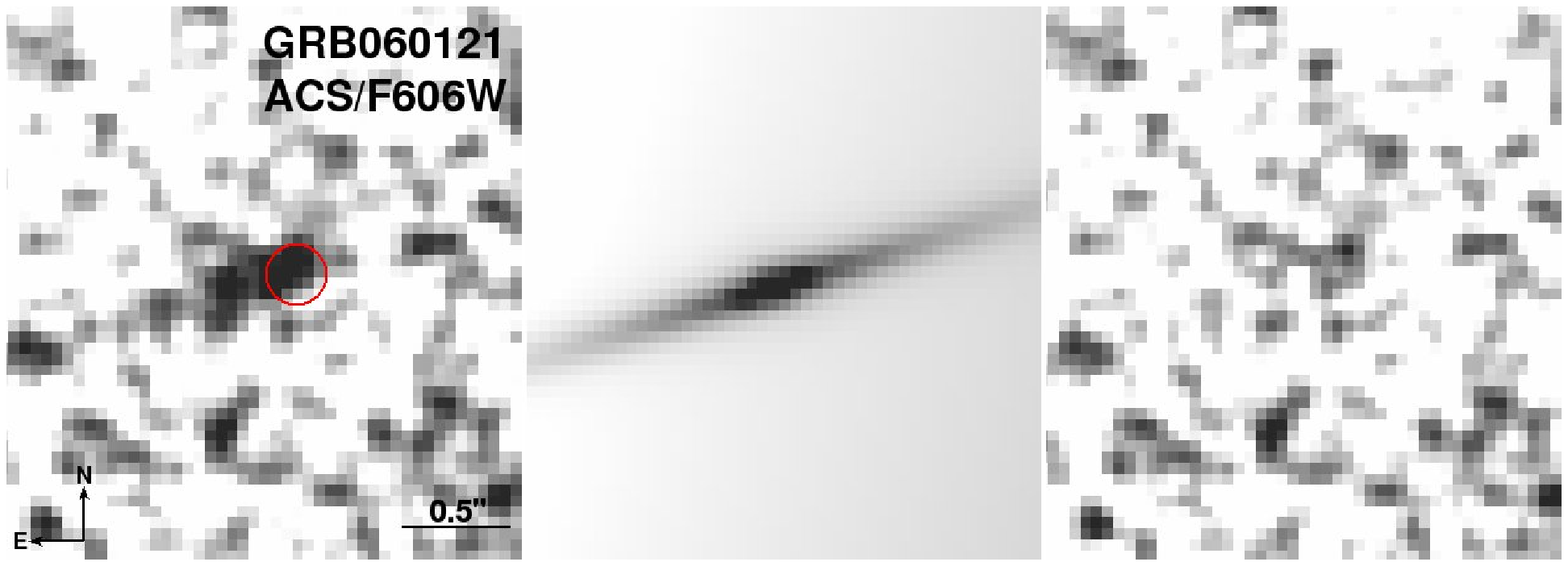}
\caption{{\it Top-left:} {\it HST}/ACS/F606W image of the host galaxy
of GRB\,060121 with a $3\sigma$ error circle representing the
afterglow position.  The image has been smoothed with a $2\times 2$
pixel Gaussian. {\it Top-center:} S\'{e}rsic model fit from {\tt
galfit} with a fixed value of $n=1$.  {\it Top-right:} Residual image.
{\it Bottom:} Same, but for S\'{e}rsic model fit from {\tt galfit}
with a fixed value of $n=4$.
\label{fig:060121}}
\end{figure}

\begin{figure}
\centering
\includegraphics[angle=0,width=5.0in]{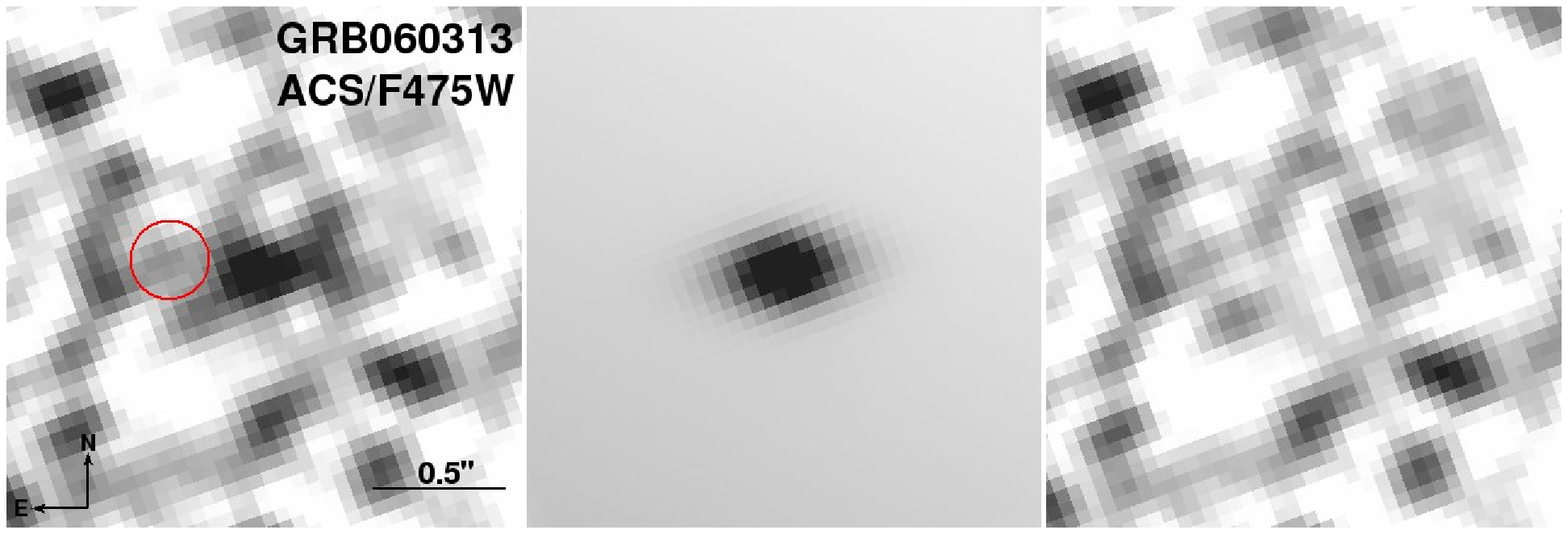}
\includegraphics[angle=0,width=5.0in]{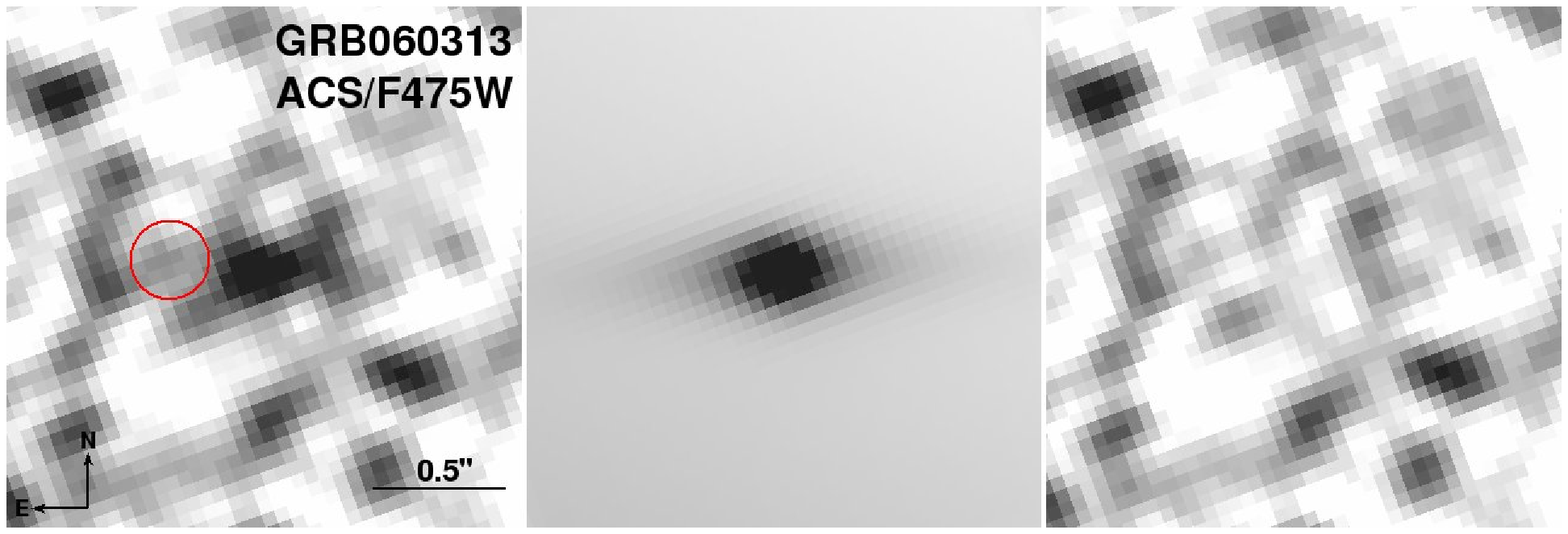}
\includegraphics[angle=0,width=5.0in]{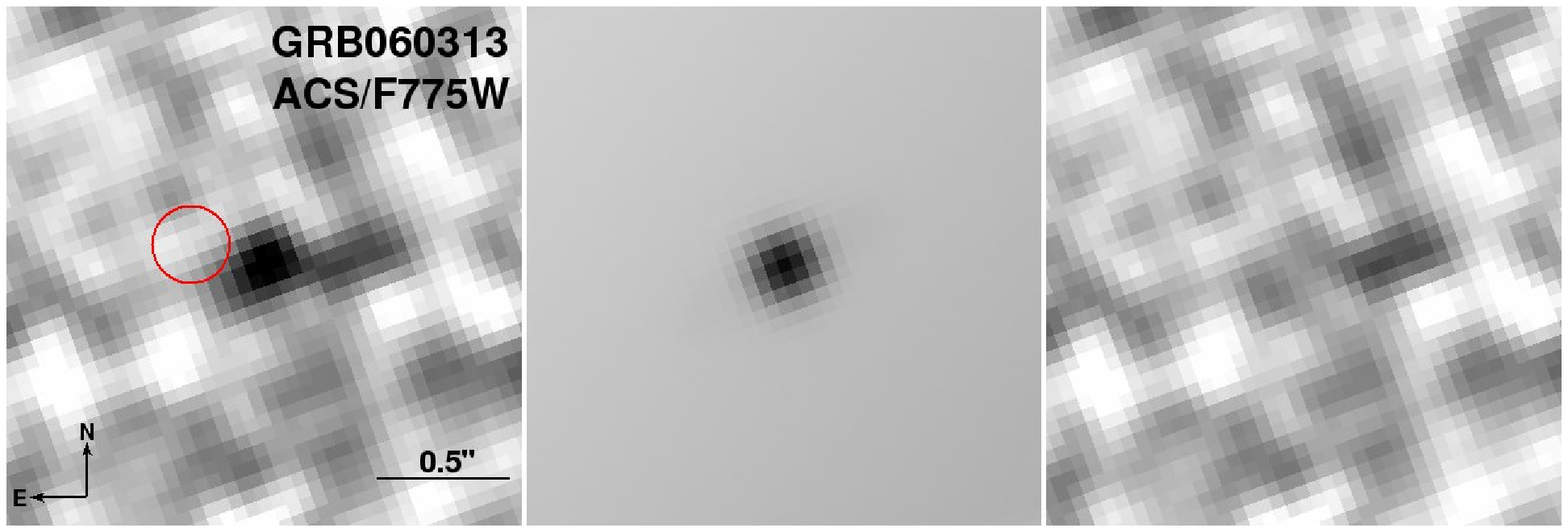}
\includegraphics[angle=0,width=5.0in]{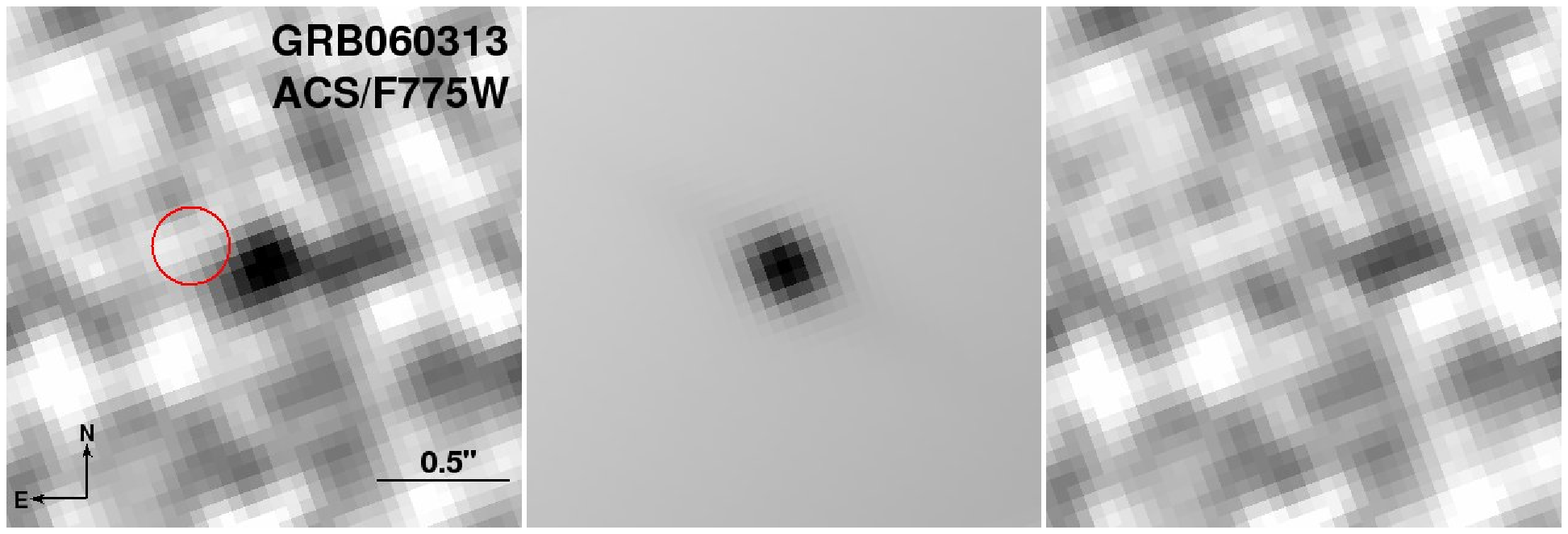}
\caption{{\it Top-left:} {\it HST}/ACS/F475W image of the host galaxy
of GRB\,060313 with a $3\sigma$ error circle representing the
afterglow position.  The image has been smoothed with a $3\times 3$
pixel Gaussian.  {\it Top-center:} S\'{e}rsic model fit from {\tt
galfit}.with a fixed value of $n=1$.  {\it Top-right:} Residual image.
{\it Second panel:} Same, but for S\'{e}rsic model fit from {\tt
galfit} with a fixed value of $n=4$.  {\it Third panel:} Same, but for
the {\it HST}/WFPC2/F775W observations with a fixed value of $n=1$.
{\it Bottom panel:} Same, but for the {\it HST}/WFPC2/F775W
observations with a fixed value of $n=4$.
\label{fig:060313}}
\end{figure}

\begin{figure}
\centering
\includegraphics[angle=0,width=5.0in]{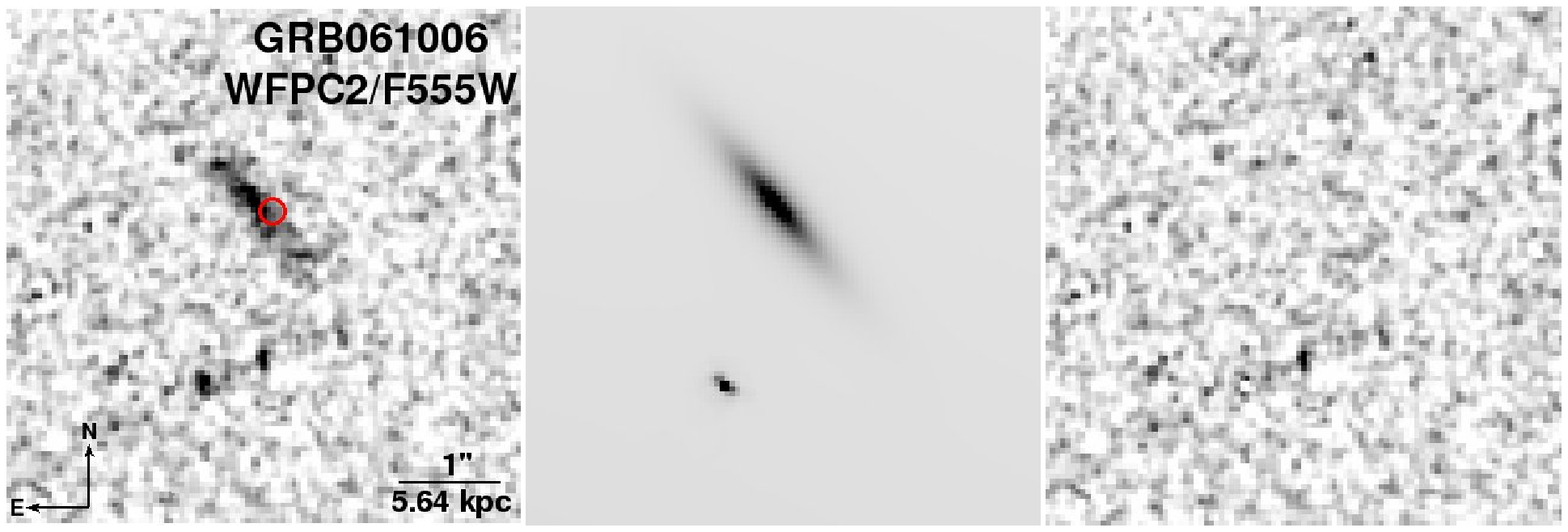}
\includegraphics[angle=0,width=5.0in]{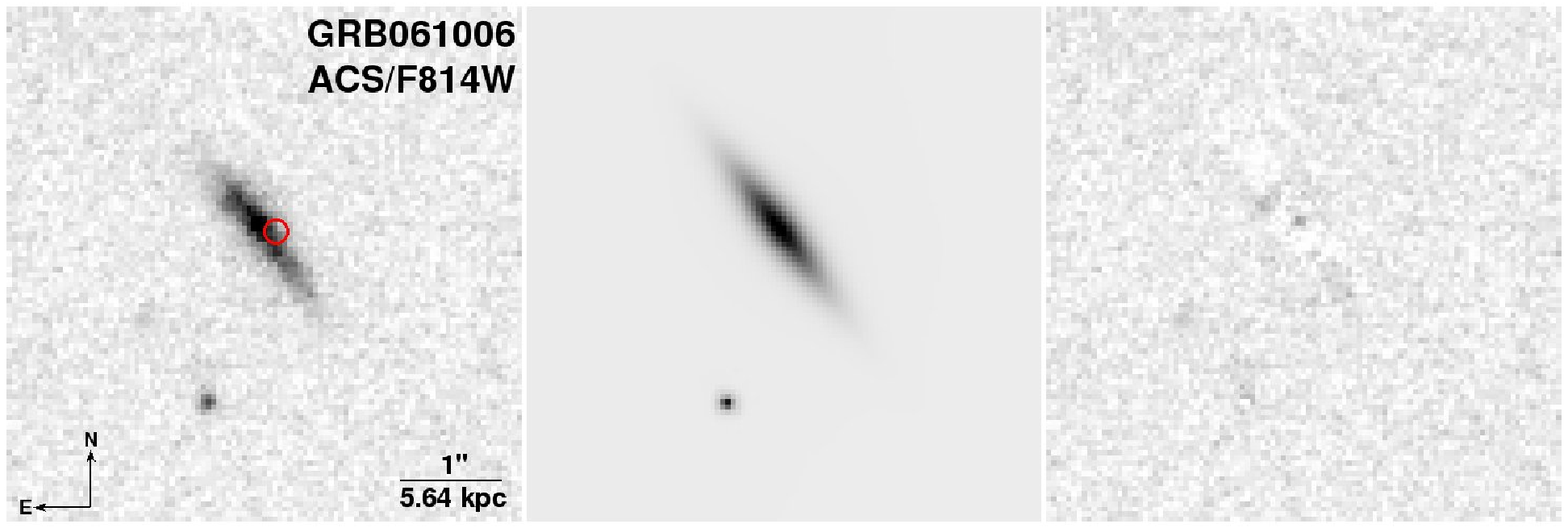}
\caption{{\it Top-left:} {\it HST}/ACS/F555W image of the host galaxy
of GRB\,061006 with a $3\sigma$ error circle representing the
afterglow position.  {\it Top-center:} S\'{e}rsic model fit from {\tt
galfit}.  {\it Top-right:} Residual image.  {\it Bottom:} Same, but
for the {\it HST}/ACS/F814W observations.
\label{fig:061006}}   
\end{figure}

We retrieved pre-processed images from the {\it HST}
archive\footnotemark\footnotetext{see http://archive.stsci.edu/hst/.}
for all available short GRBs.  We distortion-corrected and combined
the individual exposures using the IRAF task {\tt multidrizzle}
\citep{fh02,kfh+02}.  For the ACS images we used pixfrac\,=\,0.8 and
pixscale\,=\,$0.05$ arcsec pixel$^{-1}$, while for the WFPC2 images,
we used pixfrac\,=\,1.0 and pixscale\,=\,0.0498 arcsec pixel$^{-1}$,
half of the native pixel scale.  The final drizzled images,
flux-calibrated to the AB magnitude system according to the ACS and
WFPC2 zeropoints, are shown in
Figures~\ref{fig:050509b}-\ref{fig:061201}.

\subsection{Surface Brightness Profile Fitting}

We use two methods to fit the surface brightness profiles of the short
GRB host galaxies.  First, we use the {\tt galfit} software package
\citep{phi+07} to construct the best two-dimensional ellipsoidal model
of each galaxy image.  Second, we use the IRAF task {\tt ellipse} to
produce elliptical intensity isophotes and to construct
one-dimensional radial surface brightness profiles.  We further use
{\tt ellipse} to measure the integrated AB magnitude for each galaxy
(listed in Table~\ref{tab:hst}).

\subsubsection{\tt galfit}

As an input to {\tt galfit} we generate point-spread-function (PSF)
models for each instrument and filter combination using the {\tt
tinytim} software package.  We assume a constant spectrum in $F_\nu$;
the difference in the 90\% encircled energy width of the PSF for a
spectral index ranging from $-2$ to $0$ is only 1\%.  We additionally
correct for distortion in the ACS instrument, and use a sub-sampling
factor of 2 for WFPC2 to appropriately account for the reduced pixel
scale in the drizzled images.

In each observation we fit the host galaxy image with a PSF-convolved
S\'{e}rsic profile:
\begin{equation}
\Sigma(r)=\Sigma_e\,{\rm exp}\{-\kappa_n[(r/r_e)^{1/n}-1]\},
\label{eqn:sersic}
\end{equation}
where $n$ is the concentration parameter ($n=1$ is equivalent to an
exponential disk profile, while $n=4$ is the de Vaucouleurs profile),
$\kappa_n\approx 2n-1/3+4/405n+46/25515n^2$ is a constant that is
coupled to the value of $n$ \citep{cb99}), $r_e$ is the effective
radius, and $\Sigma_e$ is the effective surface brightness in flux
units.  In the subsequent discussion, tables, and figures we use
surface brightness in units of mag arcsec$^{-2}$, designated as
$\mu_e$.

In all cases we fit the host galaxies with a
single\footnotemark\footnotetext{In the case of GRBs 050509b, 050709,
060121, and 061006 we fit additional S\'{e}rsic and point-source
components to account for foreground/background objects.  These
components are not considered to be part of the host galaxy.}
S\'{e}rsic profile and allow the parameters to vary freely.  The
resulting best-fit values of $n$, $r_e$, and $\mu_e$, as well as the
integrated AB magnitudes are provided in Table~\ref{tab:morph}.  For
host galaxies that are detected at a low signal-to-noise ratio we find
that a wide range of $n$ values can account for the observed
morphology.  In these cases we fit the host galaxies with $n$ fixed at
values of 1 and 4, and provide the results of both models in
Table~\ref{tab:morph}.

The {\tt galfit} models and residual images for all instrument/filter
combinations are shown in Figures~\ref{fig:050509b}-\ref{fig:061006}.
Objects for which both $n=1$ and $n=4$ models provide an adequate fit
are shown for both cases.

\subsubsection{Radial Profiles from IRAF/{\tt ellipse}}

\begin{figure}
\centering
\includegraphics[angle=90,width=6.0in]{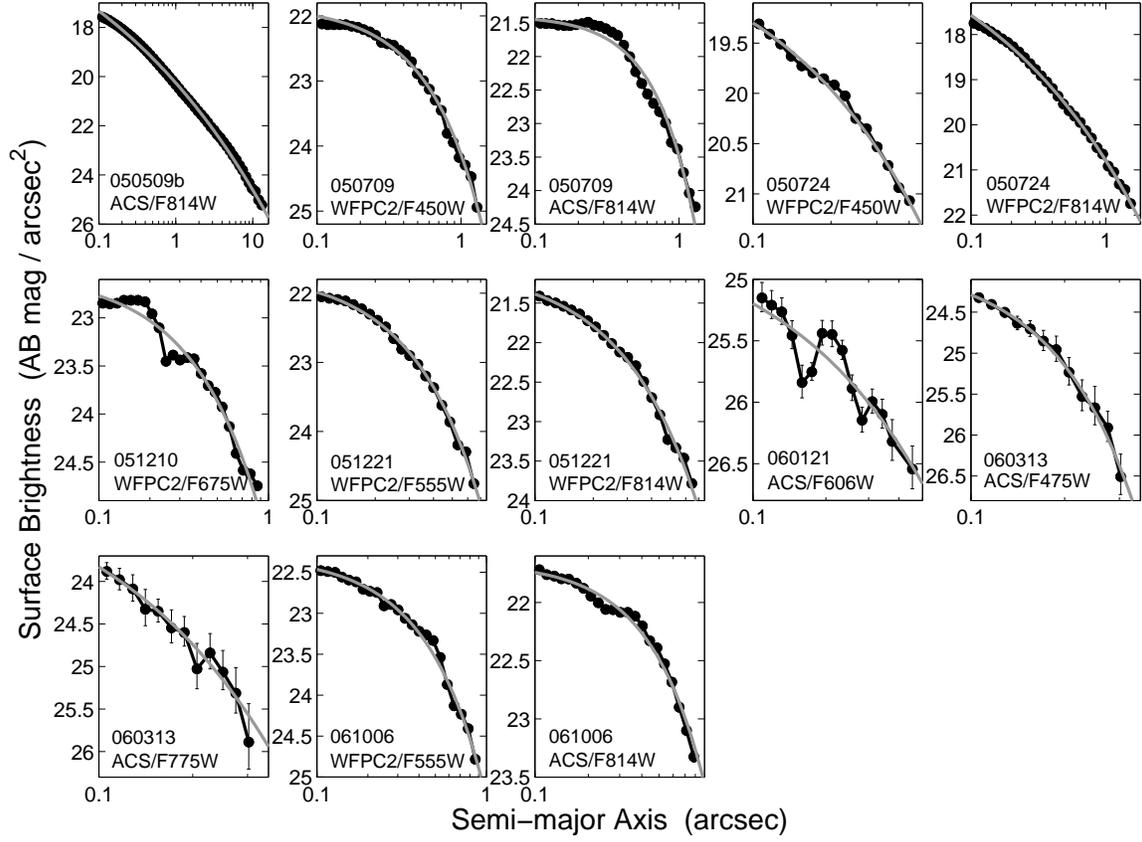}
\caption{One-dimensional radial surface brightness profiles for short
GRB host galaxies derived from IRAF/{\tt ellipse}.  The gray lines are
S\'{e}rsic model fits (Equation~\ref{eqn:sersic}) to the surface
brightness profiles.  The results of the fits are listed in
Table~\ref{tab:morph}.
\label{fig:sbfit_all}}
\end{figure}

We use {\tt ellipse} to generate elliptical isophotes for each host
galaxy, with the center and ellipticity of each isophote allowed to
vary\footnotemark\footnotetext{For the hosts of GRBs 060121 and
060313, with low signal-to-noise detections, the center and
ellipticity were held fixed throughout the fit.}.  The resulting
radial surface brightness profiles in units of AB mag arcsec$^{-2}$
are shown in Figure~\ref{fig:sbfit_all}.  We fit each profile with a
S\'{e}rsic model (Equation~\ref{eqn:sersic}) using $n$, $r_e$, and
$\mu_e$ as free parameters.  The best-fit values are listed in
Table~\ref{tab:morph}, and the resulting models are shown in
Figure~\ref{fig:sbfit_all}.  We find adequate fits in all cases,
although some host galaxies clearly exhibit radial complexity due to
irregular structure and/or an edge-on orientation.

\subsection{Astrometry}
\label{sec:astrom}

To determine the location of each short GRB relative to its host
galaxy we perform differential astrometry using optical and near-IR
images of the afterglows\footnotemark\footnotetext{Optical afterglows
have not been detected in the case of GRBs 050509b and 051210.}.  With
the exception of GRB\,050709, whose afterglow is directly detected in
{\it HST}/ACS observations, we use ground-based images from Magellan,
Gemini, and the VLT.  The astrometric tie between the afterglow and
host images is performed using point sources in common between the two
images; the source of the afterglow image and the number of
astrometric tie objects are listed in Table~\ref{tab:offsets}.  In the
case of ground-based to {\it HST} astrometry we use a range of $15-85$
common objects, with the number depending on the density of stellar
sources in the field, the depth of the images, and the field-of-view.
To determine the astrometric tie we use the IRAF astrometry routine
{\tt ccmap}.  We find that a second-order polynomial, with six free
parameters corresponding to a shift, scale, and rotation in each
coordinate, provides a robust astrometric tie in all cases.  The
resulting rms values are $\sigma_{\rm GB\rightarrow HST}=13-30$ mas
(Table~\ref{tab:offsets}).

We next consider the uncertainty in the afterglow position from each
ground-based image.  The centroiding accuracy depends on the size of
the PSF and the signal-to-noise ratio (S/N) of the afterglow
detection, $\sigma_\theta=\theta_{\rm FWHM}/({\rm S/N})$.  We
determine $\sigma_{\rm \theta,GRB}$ for each GRB using the {\tt
SExtractor} program\footnotemark\footnotetext{\tt
http://sextractor.sourceforge.net/} (Table~\ref{tab:offsets}).  In the
case of GRBs 050724 and 051221 we find that $\sigma_{\rm \theta,GRB}$
is significantly smaller than $\sigma_{\rm GB\rightarrow HST}$; for
GRBs 060121, 060313, and 061006 the two sources of uncertainty are
comparable; and for GRB\,061201 $\sigma_{\rm \theta,GRB}$ dominates.
The afterglows of GRBs 050509b, 051210, and 060502b have only been 
detected in
X-rays, with the {\it Swift} X-ray Telescope (XRT), and as a result
their positional uncertainty is $\sigma_{\rm \theta,GRB}\sim
1.7-5.8''$ (Table~\ref{tab:offsets}).  We note that the XRT positions
from the catalogs of \citet{but07} and \citet{ebp+09} exhibit relative
offsets of $5.6''$ (GRB\,050509b), $3.0''$ (GRB\,051210), and $2.9''$
(GRB\,060502b) suggesting that the true positional uncertainties 
(including systematics) are larger than their quoted statistical 
uncertainties.

The final source of uncertainty in the relative position of the GRB
and host galaxy is the centroiding accuracy of the host in the {\it
HST} images.  To determine this uncertainty we again use {\tt
SExtractor}.  The resulting values of $\sigma_{\rm \theta,gal}$ are
listed in Table~\ref{tab:offsets}.  We find that for GRBs 050724,
051221, and 061006 the host centroid uncertainty is smaller than both
$\sigma_{\rm GB\rightarrow HST}$ and $\sigma_{\rm \theta,GRB}$, while
for GRBs 060121 and 060313 $\sigma_{\rm\theta,gal}$ is comparable to
$\sigma_{\rm \theta,GRB}$.  The combined offset uncertainties are
listed in Table~\ref{tab:offsets}.

A much more accurate relative position is available for GRB\,050709
since the afterglow was detected in {\it HST}/ACS images
\citep{ffp+05}.  The limiting factor is thus $\sigma_\theta$ for the
afterglow and host.  We find from the first {\it HST}/ACS observation
(2005 July 15.6 UT) that $\sigma_{\rm\theta,GRB}=1$ mas, while
$\sigma_{\rm\theta,gal}=1.4$ mas.  In addition, we also
astrometrically tie the final epoch of ACS imaging and the WFPC2 image
to the first epoch.  Since the afterglow is no longer detected in
these images, this allows us to study the burst environment.  For the
final ACS image we find $\sigma_{\rm HST\rightarrow HST}=8$ mas, while
for the WFPC2 image we find $\sigma_{\rm HST\rightarrow HST}=14.5$
mas.  These uncertainties clearly dominate over the centroiding errors
of the afterglow and host galaxy.  We do note, however, that the
complex morphology of the host galaxy (\S\ref{sec:morph}) introduces a
systematic uncertainty in the definition of the host ``center''.  By
varying the signal-to-noise threshold in {\tt SExtractor}, we find
that the centroid of the host shifts by as much as 50 mas, which
dominates over the statistical uncertainty in the source position.

Similarly, we find a more accurate offset for GRB\,050724 from a
detection of the afterglow and host in ground-based near-IR images
\citep{bpc+05}.  The combined afterglow and host centroid uncertainty
in these images is about 6 mas, compared to a total uncertainty of
$20-25$ mas for the {\it HST} images.

\subsection{Host Light Distribution}

To determine the brightness of the GRB location relative to the host
light distribution, we follow the methodology of \citet{fls+06} and
\citet{kkp08} and calculate from each galaxy image the fraction of
total light in pixels fainter than the afterglow position.  Six bursts
have differential astrometric positions of better than 1 pixel
(050709, 050724, 051221a, 060121, 060313 and 061006;
Table~\ref{tab:offsets}).  For each image we create an intensity
histogram of a $5''\times 5''$ region centered on the host galaxy and
determine a $1\sigma$ cut-off level for the host by fitting a Gaussian
profile to the sky brightness distribution (equivalent to a
signal-to-noise ratio cut-off of 1).  We then plot the pixel flux
distribution above the appropriate cut-off level for a region
surrounding the host, and determine the fraction of light in pixels
fainter than the afterglow pixel; see Table~\ref{tab:lightfrac}.

\section{Morphological Analysis}
\label{sec:morph}

\begin{figure}
\centering
\includegraphics[angle=0,width=5.0in]{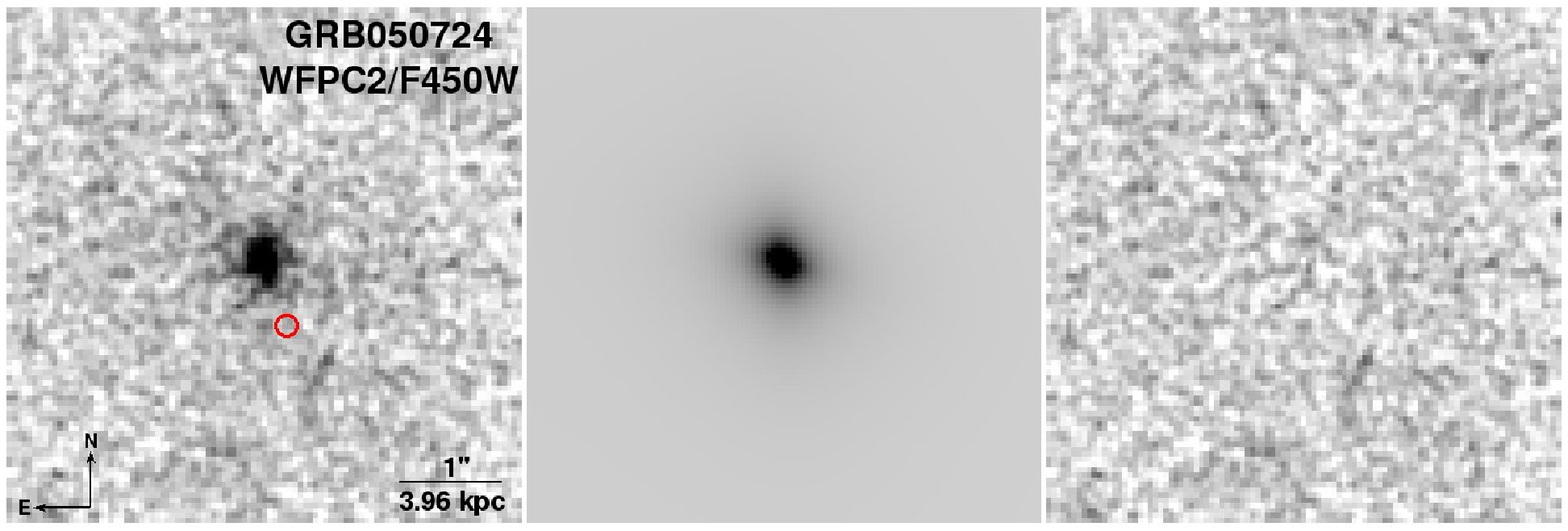}
\includegraphics[angle=0,width=5.0in]{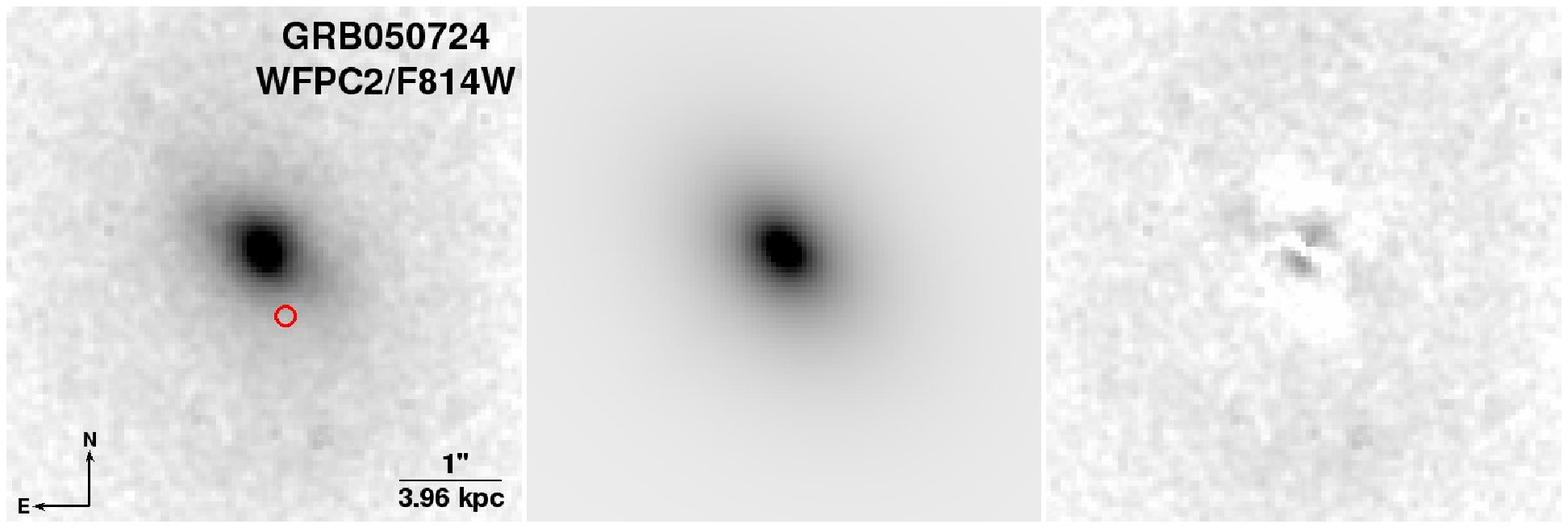}
\caption{{\it Top-left:} {\it HST}/WFPC2/F450W image of the host
galaxy of GRB\,050724 with a $5\sigma$ error circle representing the
afterglow position.  {\it Top-center:} S\'{e}rsic model fit from {\tt
galfit}.  {\it Top-right:} Residual image.  {\it Bottom:} Same, but
for the {\it HST}/WFPC2/F814W observations.
\label{fig:050724}}
\end{figure}

Using the results of the {\tt galfit} analysis and the radial surface
brightness profiles we first classify the short GRB hosts in terms of
their S\'{e}rsic $n$ values.  From the {\tt galfit} analysis we find
that three hosts (GRBs 050709, 051221a, and 061006) are best modeled
with $n\approx 1$, corresponding to an exponential disk profile, while
two hosts (GRBs 050509b and 050724) are best modeled with $n\approx 3$
and $\approx 5.6$, respectively, typical of elliptical galaxies.  We
note that GRB\,050724 possibly exhibits weak spiral structure, which
may explain the resulting value of $n\approx 3$ (see
Figure~\ref{fig:050724} and \citealt{mcd+07}), but this putative
spiral structure is clearly sub-dominant relative to the elliptical
structure.  The final three hosts (GRBs 051210, 060121, and 060313)
are equally well modeled in {\tt galfit} with a wide range of $n$
values, and we provide results for both $n=1$ and $n=4$ in
Table~\ref{tab:morph}.

\begin{figure}
\centering
\includegraphics[angle=0,width=6.0in]{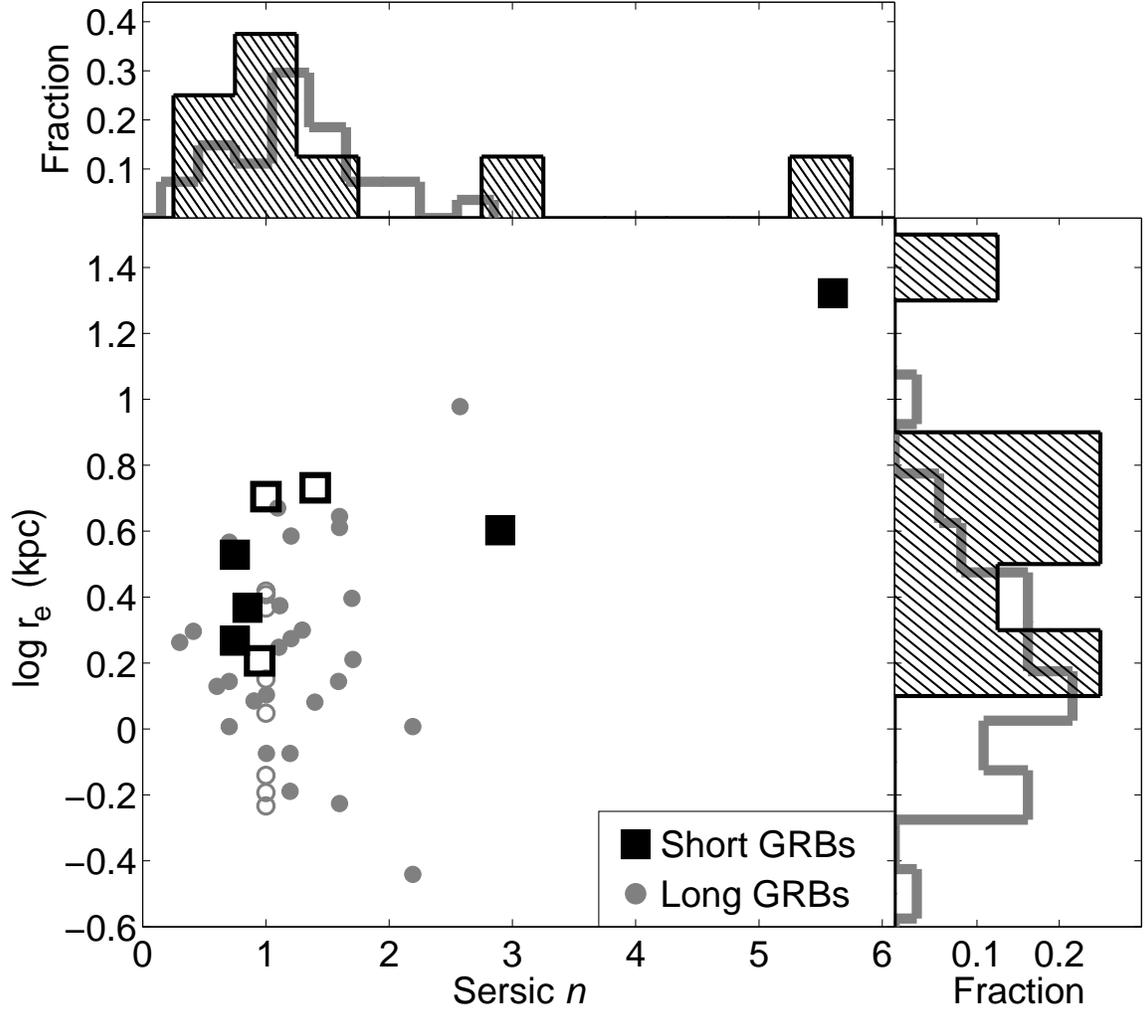}
\caption{Effective radii for the short GRB hosts with {\it HST}
observations plotted as a function of their S\'{e}rsic $n$ values.  We
use the results of the IRAF/{\tt ellipse} analysis; see
Figure~\ref{fig:sbfit_all} (open squares designate hosts for which
{\tt galfit} models with $n=1$ and $n=4$ provide an equally good fit).
Also shown are the data for long GRB hosts based on {\it HST}
observations from the sample of \citet{wbp07}.  The hosts of GRBs
050509b and 050724 have $n$ values typical of elliptical galaxies, but
the remaining hosts have a similar distribution to that of long GRBs
(i.e., a median of $n\sim 1$, or an exponential disk profile).  On the
other hand, the hosts of short GRBs are larger by about a factor of 2
than the hosts of short GRBs, in agreement with their higher
luminosities.
\label{fig:re_n}}
\end{figure}

We find identical results using S\'{e}rsic model fits to the
one-dimensional radial surface brightness profiles generated with {\tt
ellipse} (Figure~\ref{fig:sbfit_all} and Table~\ref{tab:morph}).
However, with this approach we find best-fit values of $n\sim 1$ for
the three host galaxies with ambiguous {\tt galfit} results,
suggesting that they are indeed better modeled as exponential disks.
We therefore conclude that of the eight short GRB host galaxies
studied here only two can be robustly classified as elliptical
galaxies based on their morphology.  A similar fraction was determined
independently from spectroscopic observations \citep{ber09}.  The
distribution of $n$ values is shown in Figure~\ref{fig:re_n}.

As can be seen from the {\tt galfit} results, the S\'{e}rsic models of
the two elliptical hosts exhibit significant residuals
(Figures~\ref{fig:050509b} and \ref{fig:050724}).  This is a
well-known effect for bright elliptical galaxies, which generally
require a multi-parameter power-law plus S\'{e}rsic fit that accounts
for a flatter core than expected in the de Vaucouleurs model
\citep{tea+04}.  Since we are here mainly interested in the
distribution of $n$ values and a comparison to long GRB hosts, we
retain the simple S\'{e}rsic formulation.

We also find significant residuals for a one-component S\'{e}rsic fit
of the host galaxy of GRB\,050709, which has an irregular morphology
dominated by an exponential profile (Figure~\ref{fig:050709}).  This
is the only clearly irregular galaxy in the sample.  Finally, we find
that the hosts of GRBs 051210, 060121, and 061006 exhibit significant
bulges, clearly seen in their radial surface brightness profiles
(Figure~\ref{fig:sbfit_all}).  For the host of GRB\,061006, which was
observed in two filters, the bulge component is more significant in
the F814W filter than in the F555W filter, as expected for an older
stellar population; the burst appears to coincide with this bulge
component (Figure~\ref{fig:061006}).

The {\tt galfit} and radial profile fits also yield values of the
effective radius, $r_e$, for each host galaxy.  We find a range of
$\approx 0.2-5.8''$, corresponding to physical
scales\footnotemark\footnotetext{For the faint hosts without a known
redshift (GRBs 051210, 060121, 060313, and possibly 061201) we assume
$z=1$ \citep{bfp+07}, and take advantage of the relative flatness of
the angular diameter distance as a function of redshift beyond $z\sim
0.5$.} of about $1.4-21$ kpc.  The smallest effective radius is
measured for the host of GRB\,060313, while the host of GRB\,050509b
has the largest effective radius.  The median value is $r_e\approx
3.5$ kpc.  We adopt the best-fit values from the radial surface
brightness profiles, and plot the resulting distribution, as well as
$r_e$ as a function of $n$, in Figure~\ref{fig:re_n}.

Finally, the effective surface brightness values range from $\mu_e
\approx 21$ to $\approx 27$ AB mag arcsec$^{-2}$.  The galaxy with the
highest surface brightness is the host of GRB\,050724, while the
lowest surface brightness is measured for the host of GRB\,060121.
The integrated magnitudes range from about 16.3 AB mag (GRB\,050509b)
to 26.4 AB mag (GRB\,060313).

\subsection{Comparison to Long GRB Host Galaxies}

A comprehensive morphological analysis of long GRB host galaxies using
{\it HST} observations with the STIS, WFPC2, and ACS instruments has
been carried out by \citet{cvf+05} and \citet{wbp07}.  In
Figure~\ref{fig:re_n} we compare the values of $n$ and $r_e$ measured
for long GRB hosts by \citet{wbp07} to the values measured here for
short GRB hosts.  Two clear trends emerge from this comparison.
First, all long GRB hosts have $n\lesssim 2.5$, and the median value
for the population is $\langle n\rangle\approx 1.1$ \citep{wbp07}.
Thus, all long GRB hosts are morphologically classified as exponential
disks, while 2 of the 8 short GRB hosts studied here exhibit de
Vaucouleurs profiles.  However, for the hosts with $n\lesssim 2$, the
distributions of $n$ values for both populations appear to be similar.

Second, short GRB hosts have larger effective radii, with $\langle
r_e\rangle\approx 3.5$ kpc, compared to $\langle r_e\rangle\approx
1.7$ kpc for long GRB hosts \citep{wbp07}.  A Kolmogorov-Smirnov (K-S)
test indicates that the probability that the short and long GRB hosts
are drawn from the same underlying distribution of host galaxy
effective radii is only 0.04.  If we remove from the sample GRBs
050509b and 051210 (which have only XRT positions) we find that the
K-S probability is still only 0.09.  Thus, we conclude with high
significance that short GRB host galaxies are systematically larger
than long GRB hosts, and that this result is not affected by host
associations based on XRT positions.  The larger sizes of short GRB
hosts are expected in the context of the galaxy size-luminosity
relation (e.g., \citealt{fre70}).  We recently showed that short GRB
hosts are systematically more luminous than long GRB hosts by about
$\Delta M_B\approx 1.2$ mag \citep{ber09} and therefore their sizes
are expected to be correspondingly larger.

An additional striking difference between the hosts of long and short
GRBs is the apparent dearth of interacting or irregular galaxies in
the short GRB sample.  Of the eight host galaxies studied here, we
find only one irregular galaxy (GRB\,050709) and none that appear to
be undergoing mergers.  In contrast, the fraction of long GRB hosts
with an irregular or merger/interaction morphology is about $30-60\%$
\citep{wbp07}.  The interpretation for this high merger/interaction
fraction in the long GRB sample is that such galaxies are likely
undergoing intense star formation activity triggered by the
merger/interaction process, and are therefore suitable sites for the
production of massive stars.  The lack of morphological merger
signatures in the short GRB sample indicates that if any of the hosts
have undergone significant mergers in the past, the delay time between
the merger and the production of a short GRB is $\gtrsim 10^9$ yr
(e.g., \citealt{bh92}).

\section{Offsets}
\label{sec:offsets}

We next turn to an analysis of short GRB offsets relative to the
centers of their host galaxies.  Based on the astrometric tie of the
{\it HST} host observations to ground-based afterglow observations, we
find that the projected offsets are in the range of $\approx
0.12-17.7''$ (Table~\ref{tab:offsets}).  The corresponding projected
physical offsets are about $1-64$ kpc, with a median value of about
$3$ kpc.  The largest offsets are measured for GRBs 050509b and
051210, but these are based on {\it Swift}/XRT positions with
statistical uncertainties of about 12 and 18 kpc, respectively (and
possibly larger if we consider systematic uncertainties;
\S\ref{sec:astrom}).  If we consider only the bursts with
sub-arcsecond afterglow positions we find that the largest offset is
3.7 kpc (GRB\,050709), and that the median offset for the 6 bursts is
2.2 kpc.  In the case of GRB\,061201 the host association remains
ambiguous (see Appendix~\ref{sec:app2}), but even for the nearest
detected galaxy the offset is about 14.2 kpc. The obvious caveat is
that an undetected fainter host, with $\gtrsim 25.5$ AB mag, may be
located closer to the GRB position.

To investigate the offset distribution in greater detail we supplement
the values measured here with offsets for GRBs 070724, 071227, and
090510 from ground-based observations \citep{bcf+09,gcn9353}.  In the
case of GRBs 070724 and 071227 the optical afterglows coincide with
the disks of apparent edge-on spiral galaxies \citep{bcf+09,dmc+09}.
The offsets of the three bursts are 4.8, 14.8, and 5.5 kpc,
respectively \citep{bcf+09,gcn9353}.  For GRB\,071227 we calculate the
relative offset from our Magellan/IMACS observations and find a total
($\sigma_{\rm \theta,GRB}+\sigma_{\rm \theta,gal}$) uncertainty of 65
mas, corresponding to 0.34 kpc at the redshift of the
host\footnotemark\footnotetext{This is significantly more precise than
the large uncertainty of $0.4''$ quoted by \citet{dmc+09} based on
{\it absolute} astrometry; for offset measurements differential
astrometry provides a better approach.}.

There are 7 additional events with optical afterglow identifications.
Of these bursts, two (070707 and 070714b) coincide with galaxies
\citep{pdc+08,gfl+09}, but their offsets have not been measured by the
respective authors. Based on the claimed coincidence we conservatively
estimate an offset of $\lesssim 0.5''$,
corresponding\footnotemark\footnotetext{GRB\,070714b is located at
$z=0.923$, while the redshift of GRB\,070707 is not known.  Based on
the faintness of the host, $R\approx 27.3$ mag, we assume $z=1$ to
calculate the physical offset.} to $\lesssim 4$ kpc.  Two additional
bursts (070809 and 080503) do not have coincident host galaxies to
deep limits, but the nearest galaxies are located about 6.5 and 20 kpc
from the afterglow positions,
respectively\footnotemark\footnotetext{GRB\,070809 is located 19.6 kpc
from a galaxy at $z=0.219$, and about $2.3''$ from a much fainter
galaxy, which at $z\gtrsim 1$ corresponds to 18.4 kpc.  No host is
detected at the position of GRB\,080503 in deep {\it HST}
observations, but a faint galaxy is located about $0.8''$ away, which
at $z\gtrsim 1$ corresponds to 6.5 kpc.} \citep{pbm+08,pmg+09}.  For
the final three bursts (080905, 090305, and 090426) no deep host
galaxy searches exist in the literature.

In addition to the bursts with sub-arcsecond positions, several hosts
have been identified within XRT error circles in follow-up
observations (GRBs 060801, 061210, 061217, 070429b, 070729, and
080123; \citealt{bfp+07,ber09}), but in all of these cases the offsets
are consistent with zero, or may be as large as $\sim 30$ kpc (e.g.,
\citealt{bfp+07}). For example, the offsets for GRBs 060801, 061210,
and 070429b are $19\pm 16$ kpc, $11\pm 10$ kpc, and $40\pm 48$ kpc.
We use 30 kpc as a typical upper limit on the offset for these 6
events.  We note that no follow-up observations are available in the
literature for most short GRBs with X-ray positions from 2008-2009.
Finally, about $1/4-1/3$ of all short GRBs discovered to date have
only been detected in $\gamma$-rays, precluding a unique host galaxy
association and an offset measurement.

\begin{figure}
\centering
\includegraphics[angle=0,width=6.0in]{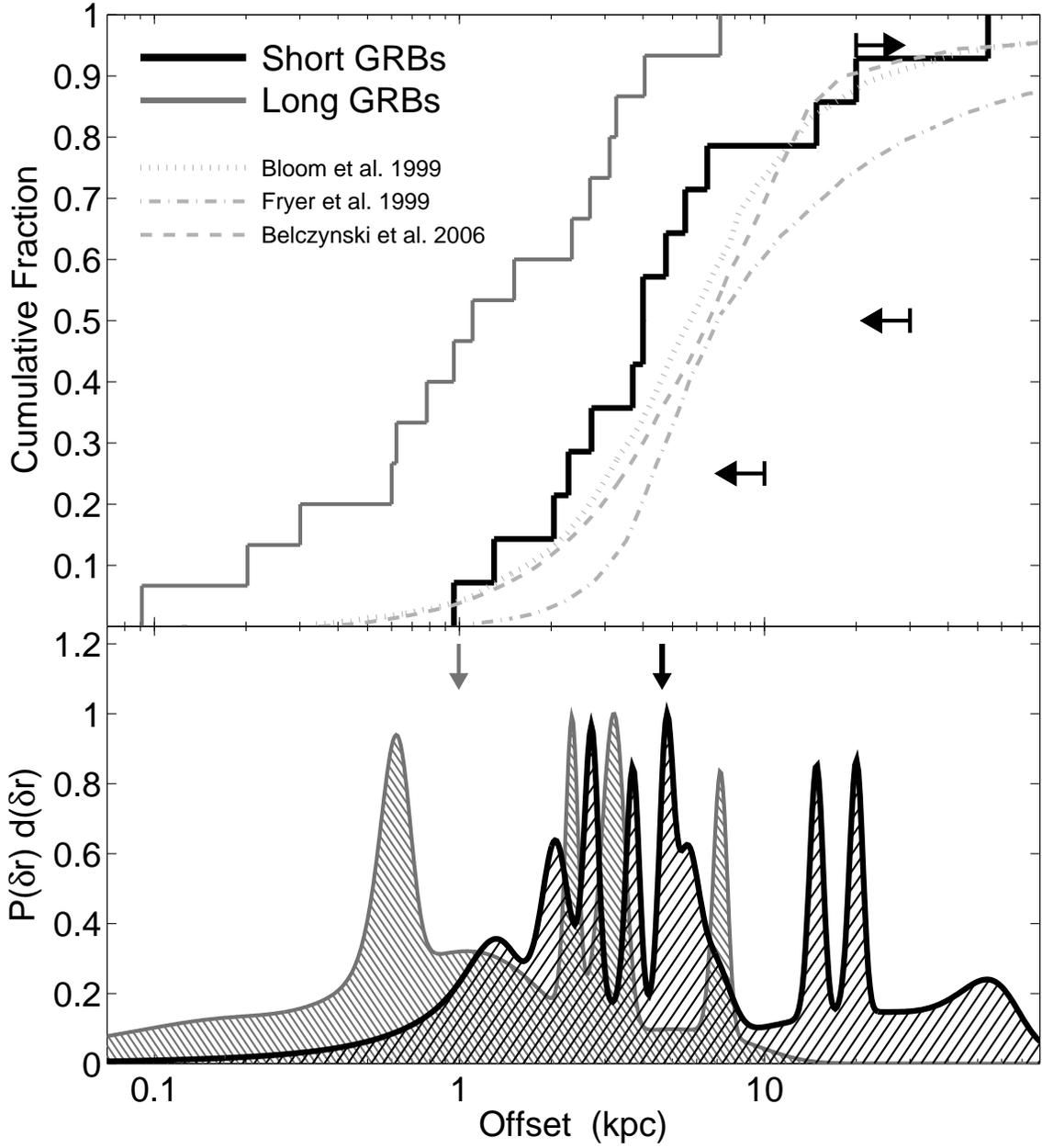}
\caption{Projected physical offsets for short GRBs (black) and long
GRBs (gray; \citealt{bkd02}).  The top panel shows a cumulative
distribution, while the bottom panel shows the differential
distribution taking into account the non-Gaussian errors on the
offsets.  The arrows in the bottom panel mark the median value for
each distribution.  The median value for short GRBs, $\approx 5$ kpc,
is about a factor of 5 times larger than for long GRBs.  The arrows in
the top panel exhibit the most robust constraints on the offset
distribution (\S\ref{sec:offsets}), taking into account the fraction
of short GRBs with only $\gamma$-ray positions, as well as short GRBs
for which hosts have been identified within XRT error circles (thereby
providing a typical range of $\sim 0-30$ kpc).  Also shown in the top
panel are predicted offset distributions for NS-NS binary mergers in
Milky Way type galaxies based on population synthesis models.  We find
good agreement between the observed distribution and models, as well
as between the robust constraints and models.
\label{fig:offsets3}}
\end{figure}

The cumulative distribution of projected physical offsets for the GRBs
with {\it HST} observations from this work, supplemented by the bursts
with offsets or limits based on optical afterglow positions (070707,
070714b, 070724, 070809, 071227, 080503, and 090510) is shown in
Figure~\ref{fig:offsets3}.  Also shown is the differential probability
distribution, $P(\delta r)d(\delta r)$, taking into account the
non-Gaussian errors on the radial offsets (see discussion in Appendix
B of \citealt{bkd02}).  We find that the median for this sample is
about 5 kpc.

As evident from the discussion above, this is not a complete offset
distribution; roughly an equal number of short GRBs have only limits
or undetermined offsets due to their detection in just the X-rays or
$\gamma$-rays\footnotemark\footnotetext{We do not consider the bursts
that lack host searches since there is no a priori reason that these
events (mainly from 2008-2009) should have a different offset
distribution compared to the existing sample from 2005-2007.}.  Taking
these events into account, our most robust inferences about the
offset distribution of short GRBs are as follows:
\begin{itemize}
\item At least $25\%$ of all short GRBs have projected physical
offsets of $\lesssim 10$ kpc.
\item At least $5\%$ of all short GRBs have projected physical offsets
of $\gtrsim 20$ kpc.  
\item At least $50\%$ of all short GRBs have projected physical
offsets of $\lesssim 30$ kpc; this value includes the upper limits for
the hosts identified within XRT error circles.
\end{itemize}
These robust constraints are shown in Figure~\ref{fig:offsets3}.

Using the observed distribution and the robust constraints outlined
above, we now provide a comparison with predicted distributions for
NS-NS binaries in Milky Way type galaxies \citep{bsp99,fwh99,bpb+06},
appropriate for the observed luminosities of short GRB host galaxies
\citep{ber09}.  We find good agreement between the observed
distribution and those predicted by \citet{bsp99} and \citet{bpb+06}.
The offset distribution of \citet{fwh99}, with a median of about 7
kpc, predicts larger offsets and therefore provides a poorer fit to
the observed distribution, which has a median of about 5 kpc.
However, all three predicted distributions accommodate the offset
constraints.  In particular, they predict about $60-75\%$ of the
offsets to be $\lesssim 10$ kpc, about $80-90\%$ to be $\lesssim 30$
kpc, and about $10-25\%$ of the offsets to be $\gtrsim 20$ kpc.  Thus,
the projected physical offsets of short GRBs are consistent with
population synthesis predictions for NS-NS binaries.  However, the
observations are also consistent with partial contribution from other
progenitor systems with no expected progenitor kicks, such as WD-WD
binaries.

\subsection{Host-Normalized Offsets}

\begin{figure}
\centering
\includegraphics[angle=0,width=6.0in]{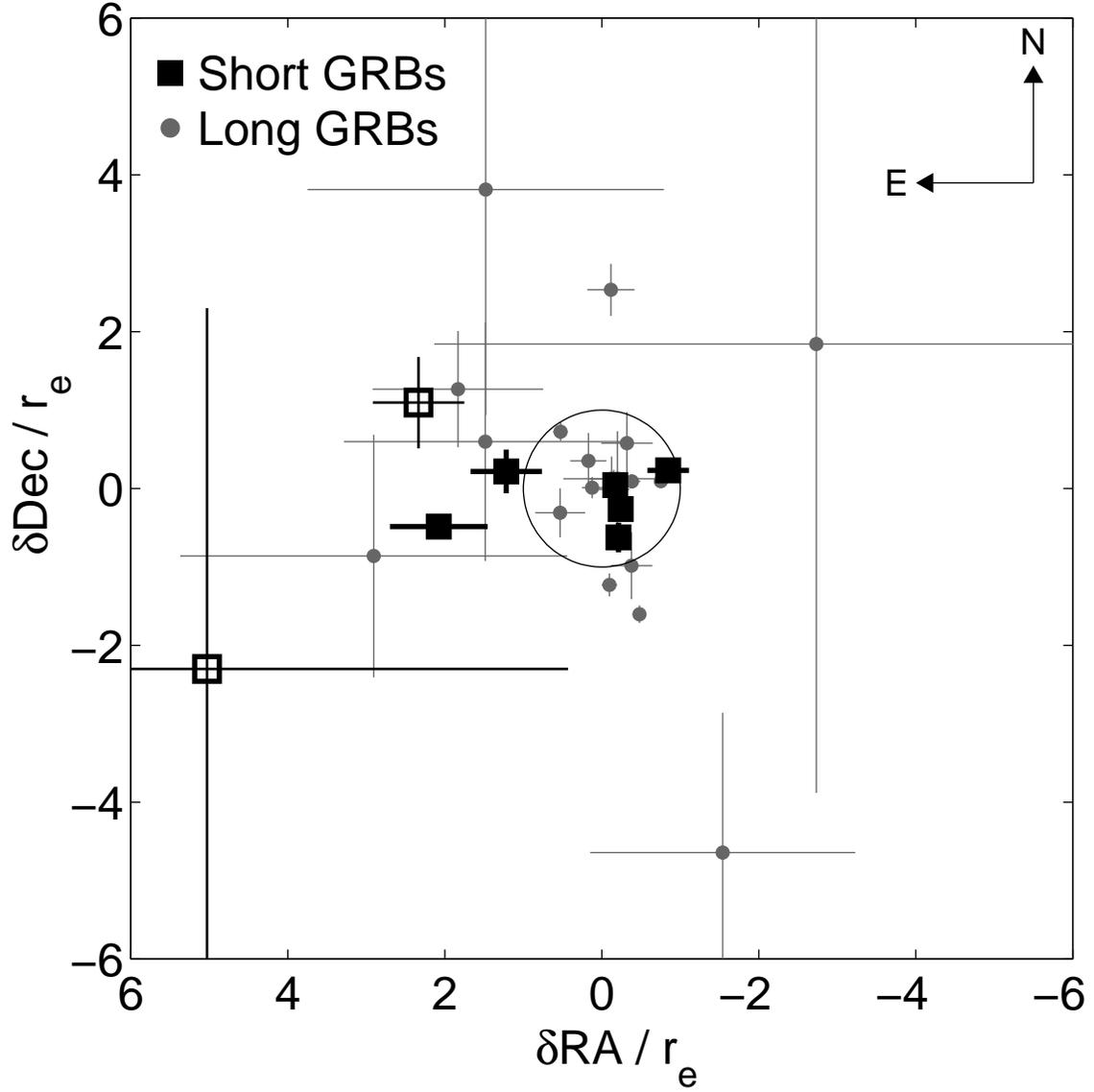}
\caption{Host-normalized offsets in right ascension and declination 
for the short GRBs in our {\it HST} sample (black; open symbols mark
the GRBs with X-ray positions, 050509b and 051210).  Also shown are
the offsets for long GRBs from the sample of \citet{bkd02}.  The circle 
marks an offset of 1 $r_e$.  About half of all long GRBs have offsets
of $\lesssim 1$ $r_e$, and we find a similar result for short GRBs.
\label{fig:offsets1}}
\end{figure}

To compare the offsets in a more uniform manner, we normalize the
measured values by $r_e$ for each host galaxy.  We use the $r_e$
values measured from the one-dimensional radial surface brightness
profiles from {\tt ellipse} (see Figure~\ref{fig:offsets1} and
Table~\ref{tab:offsets}) and find values ranging from about 0.2 $r_e$
for GRB\,060121 to $6.7\pm 2.7$ $r_e$ for the \citet{ebp+09} XRT
position of GRB\,051210.  The \citet{but07} position for GRB\,051210,
however, leads to an offset of $4.65\pm 4.60$ $r_e$, consistent with a
negligible offset.  For the subset of 6 bursts with optical afterglow
positions and secure host associations, 4 are located within 1 $r_e$,
while the remaining 2 bursts are located at about 2 $r_e$
(Figure~\ref{fig:offsets1}).  GRB\,050509b, which has the largest
physical offset, has a normalized offset of $2-3$ $r_e$, depending on
which XRT position is used.  Thus, with the exception of the ambiguous
case of GRB\,061201, we find that all of the available offsets are
consistent with $\lesssim 2$ $r_e$.  The large additional sample of
physical offsets that we used above cannot be easily translated to
host-normalized offsets at the present since none of the hosts have
been observed with {\it HST}, thereby precluding a robust
morphological analysis.  This provides an impetus for future {\it HST}
observations.

The differential probability distribution of host-normalized offsets
for our {\it HST} sample, taking into account the non-Gaussian errors,
is shown in Figure~\ref{fig:offsets2}.  We find that the median value
for all 8 bursts is $\approx 1$ $r_e$.  Moreover, $\lesssim 20\%$ of
the probability distribution is at large offsets of $\gtrsim 2.5$
$r_e$.

\subsection{Comparison to Long GRB Host Galaxies}

\begin{figure}
\centering
\includegraphics[angle=0,width=6.0in]{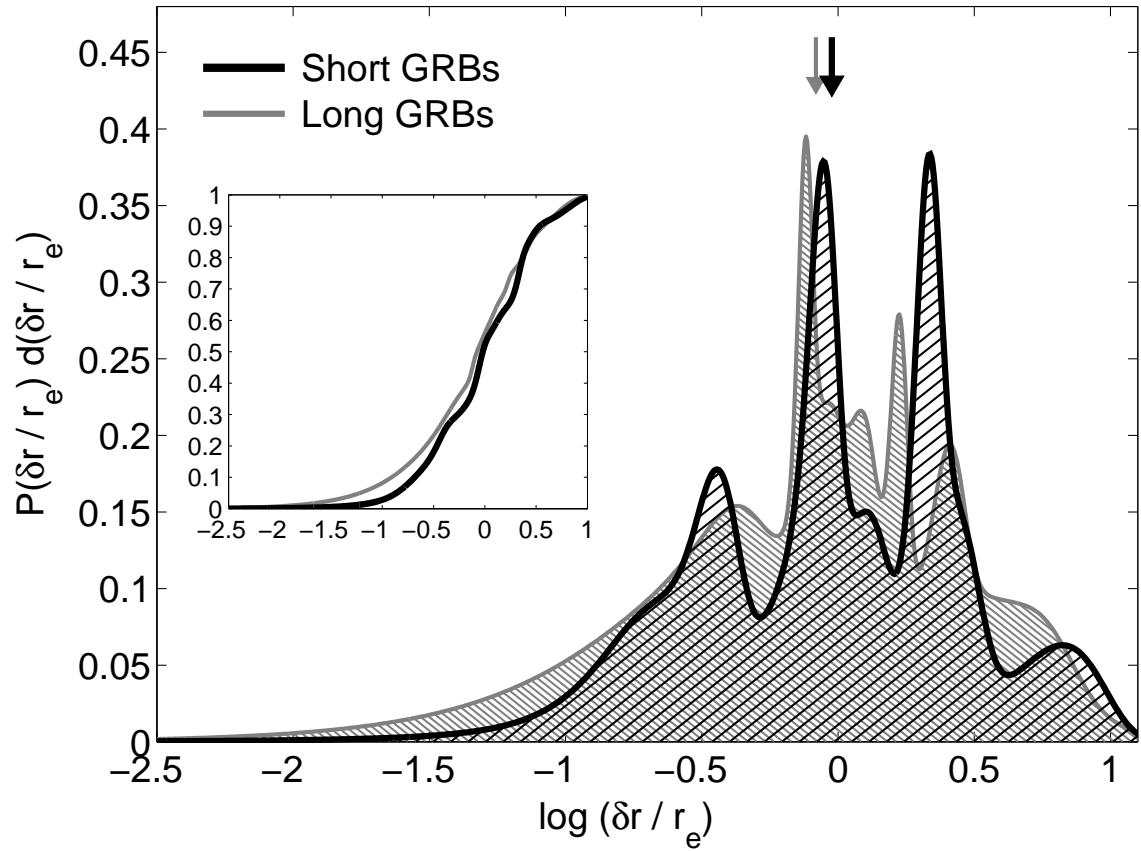}
\caption{Probability distributions of the host-normalized offsets of
short GRBs (black) and long GRBs (gray; \citealt{bkd02}).  For each
burst we include the host-normalized offset taking into account the
non-Gaussian errors.  The arrows mark the median value of each
distribution, and the inset shows the cumulative distribution.
\label{fig:offsets2}}
\end{figure}

We compare our observed short GRB offsets with those of long GRBs from
the sample of \citet{bkd02} in
Figures~\ref{fig:offsets3}-\ref{fig:offsets2}.  The offset
distribution of long GRBs has been used to argue for a massive star
progenitor population, and against NS-NS binaries \citep{bkd02}.  The
offset distribution for short GRBs is clearly shifted to larger
physical scales.  In particular, the median offset for the long GRBs
is 1.1 kpc, about a factor of 5 times smaller than the median value
for short GRBs.  Similarly, no long GRB offsets are larger than about
7 kpc, whereas at least some short GRBs appear to have offsets in
excess of 15 kpc.

However, the distinction between the two offset distributions is
significantly reduced when we take into account the systematically
larger sizes of short GRB host galaxies (Figure~\ref{fig:re_n}).  The
median normalized offset for long GRBs is about 0.8 $r_e$, compared to
about 1 $r_e$ for short GRBs.  Similarly, $20\%$ of the long GRB
cumulative distribution has offsets of $\gtrsim 2.5$ $r_e$, identical
to the statistics for the short GRB offsets.  Indeed, as can be see
from Figure~\ref{fig:offsets2}, the cumulative host-normalized offset
distributions for long GRBs and short GRBs with {\it HST} observations
are nearly identical.

In the context of NS-NS binary progenitors, the close similarity in
the normalized offset distributions can be interpreted to mean that
most systems likely remain bound to their hosts (rather than ejected
into the intergalactic medium), and/or have a relatively short delay
time.  These conclusions are tentative due to the small number of
events with host-normalized offsets, but they can be further tested
with future {\it HST} observations.

\section{Light Distribution Analysis}
\label{sec:light}

In addition to the offset analysis in the previous section, we study
the local environments of short GRBs using a comparison of their local
brightness to the host light distribution.  This approach is
advantageous because it is independent of galaxy morphology, and does
not suffer from ambiguity in the definition of the host center (see
\citealt{fls+06}).  We note that for the overall regular morphology of
short GRB hosts the definition of the host center is generally robust,
unlike in the case of long GRBs \citep{fls+06,wbp07}.  On the other
hand, this approach has the downside that it requires precise
pixel-scale positional accuracy.  In our sample, this is the case for
only 6 short bursts.

\begin{figure}
%\centerline{\psfig{file=fig15a.ps,width=3.5in,angle=0}
%\hspace{0.2in}\psfig{file=fig15b.ps,width=3.5in,angle=0}}
%\includegraphics[angle=0,width=3in]{fig15a.ps}\hspace{0.2in}
%\includegraphics[angle=0,width=3in]{fig15b.ps}
\includegraphics[angle=0,width=6.0in]{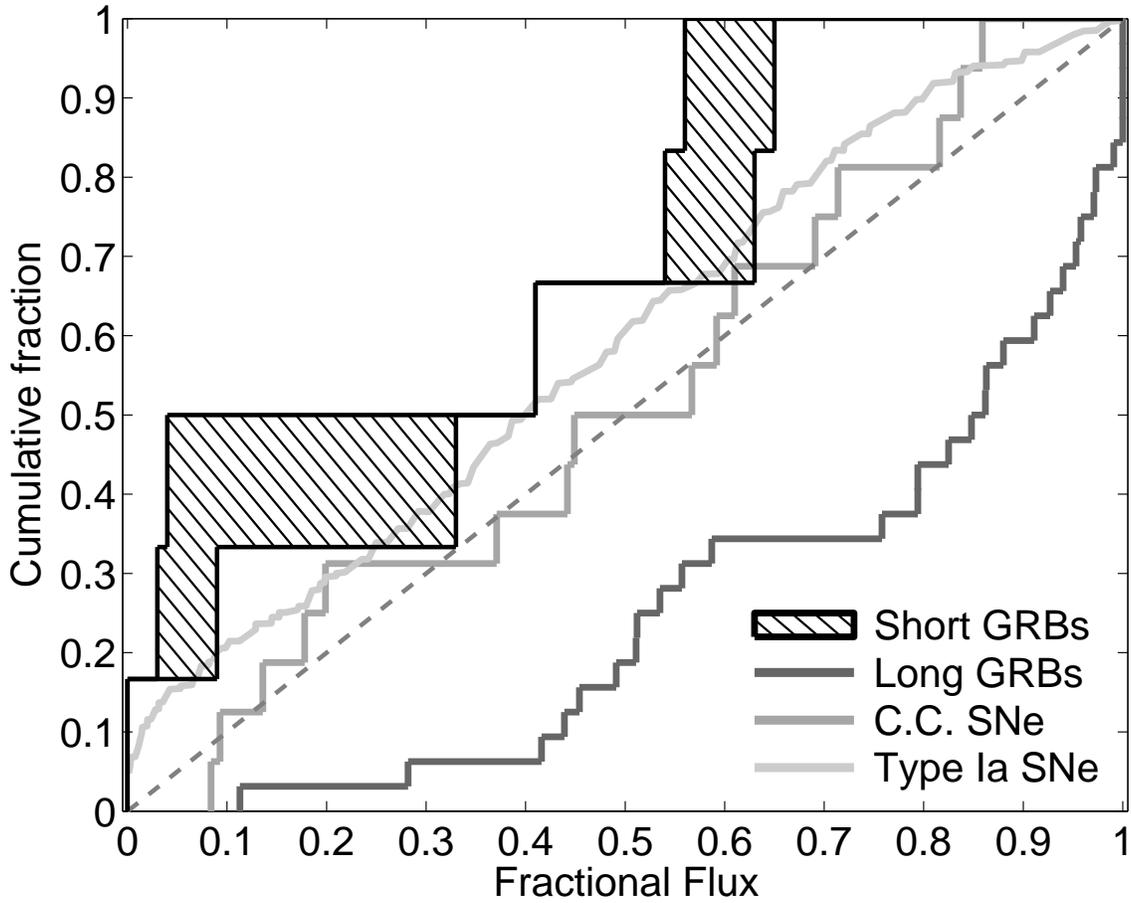}
\caption{Cumulative distribution of fractional flux at the location of
short GRBs relative to their host light.  For each burst we measure
the fraction of host light in pixels fainter than the GRB pixel
location.  The shaded area is defined by the results for the two
available filters for each short GRB.  Also shown are data for long
GRBs (dark gray line) and for core-collapse and Type Ia SNe (light
gray lines) from \citet{fls+06} and \citet{kkp08}.  The dashed line
marks the expected distribution for objects which track their host
light distribution.  Short GRBs appear to under-represent their host
light, while long GRBs tend to be concentrated in the brightest
regions of their hosts \citep{fls+06}.
\label{fig:lightfraction}}
\end{figure}

The fraction of host light in pixels fainter than the afterglow pixel
brightness for each host/filter combination is summarized in
Table~\ref{tab:lightfrac}.  The cumulative light distribution
histogram is shown in Figure~\ref{fig:lightfraction}.  The shaded
histogram represents the range defined by the dual filters for 5 of
the 6 bursts.  We find that the upper bound of the distribution is
defined by the blue filters, indicating that short GRBs trace the
rest-frame optical light of their hosts better than the rest-frame
ultraviolet.  This indicates that short GRB progenitors are likely to
be associated with a relatively old stellar population, rather than a
young and UV bright population.

The overall distribution has a median value of $\approx 0.1-0.4$
(red); namely, only in about one-quarter of the cases, $50\%$ of the
host light is in pixels fainter than at the GRB location.  Thus, the
overall distribution of short GRB locations under-represents the host
galaxies' light distribution.  This is also true in comparison to
the distribution for core-collapse SNe, which appear to track their
host light \citep{fls+06}, and even Type Ia SNe, which have a median
of about 0.4 \citep{kkp08}.  Thus, the progenitors of short GRBs
appear to be more diffusely distributed than Type Ia SN progenitors.

\subsection{Comparison to Long GRB Host Galaxies}

An extensive analysis of the brightness distribution at the location
of long GRBs has been carried out by \citet{fls+06}.  These authors
find that long GRBs are more concentrated on the brightest regions of
their hosts than expected from the light distribution of each host.
In particular, they conclude that the probability distribution of GRB
positions is roughly proportional to the surface brightness squared.
As can be seen from Figure~\ref{fig:lightfraction}, short GRBs have a
significantly more diffuse distribution relative to the host light
than long GRBs.  In particular, for the latter, the median light
fraction is about 0.85 compared to about $0.25\pm 0.15$ for the short
GRBs.

\section{Discussion and Implications}
\label{sec:disc}

Our extensive analysis of short GRB host galaxy morphologies and the
burst local environments has important implications for the progenitor
population.  We address in particular the popular NS-NS merger model,
as well as delayed magnetar formation via WD-WD mergers or WD
accretion-induced collapse \citep{mqt08}.

\subsection{Morphology}

From the morphological analysis we find continued evidence that the
bulk of short GRB host galaxies ($\sim 3/4$) are late-type galaxies,
in agreement with results from spectroscopic observations
\citep{ber09}.  Moreover, as demonstrated by the systematic
differences in luminosity, star formation rates, and metallicities
between the star forming hosts of long and short GRBs \citep{ber09},
we find here that short GRB hosts are systematically larger than long
GRB hosts.  These results indicate that the progenitors of the two GRB
classes select different environments.  The higher luminosities,
larger sizes, and lower specific star formation rates of short GRB
hosts suggest that their rate of occurrence is tied to galactic mass
rather than to star formation activity.  This result is in broad
agreement with old progenitor populations such as NS-NS, NS-BH, or
WD-WD binaries, but it indicates that the bulk of short GRB
progenitors are not young magnetars.  This conclusion is also
supported by the dearth of merger signatures, which point to delays of
$\gtrsim 10^9$ yr relative to any merger-triggered star formation
episodes.

\subsection{Offsets}

The differential offsets measured here from the {\it HST} observations
provide the most precise values to date for short GRBs, with a total
uncertainty of only $\sim 10-60$ mas, corresponding to $\sim 30-500$
pc.  We find that none of the offsets are smaller than $\sim 1$ kpc,
while this is the median offset for long GRBs.  On the other hand, a
substantial fraction of the {\it measured} offsets are only a few kpc.
The median offset for the {\it HST} observations supplemented by
ground-based data is about 5 kpc (Figure~\ref{fig:offsets3}), about 5
times larger than for long GRBs.

As discussed in detail in \S\ref{sec:offsets}, the observed offset
distribution is incomplete.  About $1/4-1/3$ of all short GRBs have
only $\gamma$-ray positions ($\sim 1-3'$), and a similar fraction have
only XRT positions, which generally lead to a range of offsets of
$\sim 0-30$ kpc.  Taking these limitations into account we find that
the most robust constraints on the offset distribution are that
$\gtrsim 25\%$ of all short GRBs have offsets of $\lesssim 10$ kpc,
and that $\gtrsim 5\%$ have offsets of $\gtrsim 20$ kpc.  Both the
observed offset distribution and these constraints are in good
agreement with predictions for the offset distribution of NS-NS
binaries in Milky Way type galaxies \citep{bsp99,fwh99,bpb+06}.
However, at the present they cannot rule out at least a partial
contribution from other progenitor systems such as delayed magnetar
formation and even young magnetar flare.  The apparent existence of
large offsets in the sample suggests that these latter models are not
likely to account for {\it all} short GRBs.

In contrast to the larger physical offsets of short GRBs, we find that
the distribution of offsets normalized to the host galaxy effective
radii exhibits much better agreement between long and short GRBs
(Figure~\ref{fig:offsets2}).  The medians of the two distributions are
similar (1 versus 0.8 $r_e$ for short and long GRBs, respectively),
and both populations have $\approx 20\%$ probability for offsets of
$\gtrsim 2.5$ $r_e$.  Naturally, due to the lack of {\it HST}
observations for short GRBs from 2007-2009, the sample for which this
analysis is possible is smaller than the sample with physical offsets.
Thus, further {\it HST} observations of existing and future short GRB
hosts are essential in order to determine whether the broad similarity
in host-normalized offsets is robust.  We stress that in the context
of comparing short GRBs with various progenitor populations and with
long GRBs, host-normalized offsets are the more relevant quantity.
For example, the host-normalized distributions for massive stars in
small and large galaxies will be similar even though their physical
offsets will differ.  We stress that population synthesis modelers
should include an appropriate distribution of host galaxy sizes, and
thereby provide predictions for host-normalized offsets.

In the context of implications for the progenitor population, a recent
study of short GRB physical offsets by \citet{tko+08} led these
authors to claim that short GRBs with extended X-ray emission have
systematically smaller offsets, possibly due to a systematic
difference in the progenitors.  Our {\it HST} sample includes three
short GRBs with strong extended emission (050709, 050724, and 061006),
and one GRB (060121) with possible extended emission ($4.5\sigma$
significance; \citealt{dls+06}).  The physical offsets of these bursts
are about 3.7, 2.7, 1.3, and 1 kpc, respectively, leading to a mean
offset of about 2.2 kpc.  The physical offsets of the bursts without
extended emission, but with precise afterglow positions (051221,
060313, and 061201) are 2.0, 2.3, and 14.2 or 32.5 kpc, respectively.
The two events with no extended emission and with XRT positions
(050509b and 051210) have offsets of about $54\pm 12$ and $28\pm 23$
kpc, respectively.  If we include the ground-based sample with optical
afterglow positions (see \S\ref{sec:offsets}), we find that the bursts
with apparent extended emission (070714b, 071227, 080513, and 090510;
\citealt{gcn6623,gcn7156,gcn9337,pmg+09}) have offsets of $\lesssim
4$, 14.8, $\sim 20$, and $\sim 5.5$ kpc, while the bursts without
extended emission (070724 and 070809) have offsets of 4.8 and $\sim
6.5$ kpc.  Thus, based on the sample of events with sub-arcsecond
positions we find that 6/8 bursts with extended emission have offsets
of $\lesssim 5$ kpc and 2/8 have likely offsets of $\sim 15-20$ kpc.
In the sample without extended emission we find that 4/5 have offsets
of $\lesssim 6$ kpc and 1/5 has a likely offset of $\sim 14-32$ kpc.
Thus, we conclude that there is no significant difference in the two
offset distributions.

The inclusion of events with only XRT positions does not change this
conclusion.  In particular, of the subset with no extended emission
only GRB\,050509b is likely to have a significant offset, while GRBs
051210, 060801, and 070429b have offsets ($28\pm 23$, $19\pm 16$, and
$40\pm 48$ kpc, respectively) that are consistent with zero.
Similarly, GRB\,061210 with extended emission has an offset of $11\pm
10$ kpc.  An examination of the sample of \citet{tko+08} reveals that
their claim that short GRBs without extended emission have
systematically larger offsets rests on four events in particular: GRBs
050509b, 060502b, 061217, and 061201.  As noted above, GRBs 050509b
and 061201 indeed appear to have substantial
offsets\footnotemark\footnotetext{We note that \citet{tko+08} assume
that the host of GRB\,061201 is the galaxy at $z=0.111$ at an offset
of about 32 kpc.  However, as we have shown here based on the {\it
HST} observations (Appendix~\ref{sec:app2}), there is a fainter
potential host at a smaller offset.}, but so do GRBs 071227 and 080503
with extended emission and offsets of about $15-20$ kpc.  Next, the
large offset for GRB\,060502b relies on its claimed association with
an elliptical galaxy $70\pm 16$ kpc from the XRT position
\citep{bpc+07}.  However, the XRT error circle contains additional
galaxies with negligible offsets \citep{bfp+07}.  Finally, we note
that the offset for GRB\,061217 is unreliable due to a substantial
discrepancy of about $33''$ in the XRT positions from \citet{but07}
and \citet{ebp+09}.  A continued investigation of the difference
between short GRBs with and without extended emission will greatly
benefit from the use of host-normalized offsets.

\subsection{Light Distribution}

In addition to projected offsets relative to the host center, we find
that the locations of the short GRBs with {\it HST} imaging and
sub-arcsecond positions are more diffusely distributed relative to
their host light than long GRBs.  In particular, we find that short
GRB positions under-represent their host light, even in comparison to
core-collapse and Type Ia SNe.  This result is likely an upper limit
on the brightness of short GRB locations since only the subset of
events with optical afterglow positions can be studied with this
approach.  Thus, short GRBs arise from a population of events with a
more diffuse distribution than massive stars and Type Ia SN
progenitors.  This result also indicates that the bulk of the
progenitors of long and short GRBs cannot both be magnetars.

There are currently 10 known short GRBs with optical afterglows for
which {\it HST} observations will enable a similar analysis.  This is
twice the number of the current sample, and we can therefore make
significant progress in understanding the relation of short GRB
environments to the overall distribution of light in their host
galaxies with future observations.

\section{Conclusions}
\label{sec:conc}

We presented the first comprehensive analysis of short GRB {\it HST}
observations, and used these data to extract the morphological
properties of the host galaxies, the projected physical and
host-normalized GRB offsets, and the brightness at the location of the
bursts relative to the overall light distribution of their hosts.  The
main conclusions of our analysis are as follows:
\begin{itemize}
\item The majority of short GRB hosts are consistent with or have
exponential surface brightness profiles, typical of late-type
galaxies.  This conclusion is in good agreement with results from
spectroscopic observations that reveal star formation activity in
$\sim 3/4$ of short GRB hosts \citep{ber09}.
\item The host galaxies of short GRBs are on average larger by about a
factor of 2 than the hosts of long GRBs.
\item The observed short GRB offset distribution extends from $\sim 1$
to 50 kpc, with a median of about 5 kpc.  Including the short GRBs
with only $\gamma$-ray or X-ray positions, we find that $\gtrsim 25\%$
of all events have offsets of $\lesssim 10$ kpc, and $\gtrsim 5\%$
have offsets of $\gtrsim 20$ kpc.  A additional, though softer, limit
is that $\gtrsim 50\%$ have offsets of $\lesssim 30$ kpc.
\item The observed physical offset distribution and the robust
constraints compare favorably with the predicted distribution for
NS-NS binaries.  However, they do not rule out at least a partial
contribution from other progenitors systems such as WD-WD binaries.
\item We find no convincing evidence that short GRBs with extended
emission have smaller physical offsets than those without extended
emission.  In both sub-samples we find examples of both small offsets
($\sim {\rm few}$ kpc) and possibly large offsets (tens of kpc).
\item The distribution of host-normalized offsets for the subset of
short GRBs with {\it HST} observations is nearly identical to that of
long GRBs.  This is due to the systematically larger size of short GRB
hosts, and indicates that a comparison with long GRBs and progenitor
models will benefit from the use of host-normalized (rather than
physical) offsets.
\item The locations of short GRBs with sub-arcsecond positions and
{\it HST} imaging under-represent the overall light distribution of
their hosts, but less so in the red.  This result differs
substantially from long GRBs, core-collapse SNe, and even Type Ia SNe.
\end{itemize}

The results derived in this paper are based mainly on a small sample
of short GRBs (9 events) from 2005-2006.  Seven of these objects have
precise positions based on optical afterglow detections.  Ten
additional events with precise afterglow positions, and a similar
number with XRT positions (some of which with identified hosts), are
now available for a similar study.  It is essential to observe this
existing sample, as well as new events from {\it Swift} and {\it
Fermi}, with the refurbished {\it HST} using the ACS and WFPC3
instruments.  In conjunction with constraints on the progenitor
population from the redshift distribution \citep{bfp+07} and
spectroscopic studies of the host galaxies \citep{ber09}, the
continued use of high angular resolution imaging will provide crucial
insight into the nature of the progenitors and the potential for
multiple populations.

\acknowledgements

\appendix
\section{The Host Galaxy of GRB\,060313}
\label{sec:app1}

We present the first host galaxy association for GRB\,060313, using
{\it HST}/ACS observations in the F475W and F775W filters.  The offset
between the GRB position (determined from Gemini-South observations;
\citealt{bfp+07}) and the galaxy center is about $0.32''$
(Table~\ref{tab:offsets}).  The galaxy brightness is $m({\rm F475W})=
26.4$ AB mag and $m({\rm F775W})=25.6$ AB mag (Table~\ref{tab:hst}).
The probability of chance coincidence at this offset and galaxy
brightness is only about $3\times 10^{-3}$ \citep{bsk+06}.  We thus
conclude that this galaxy is the likely host of GRB\,060313.

\section{Possible Host Galaxies of GRB\,061201}
\label{sec:app2}

\begin{figure}
\centering
\includegraphics[angle=0,width=5.0in]{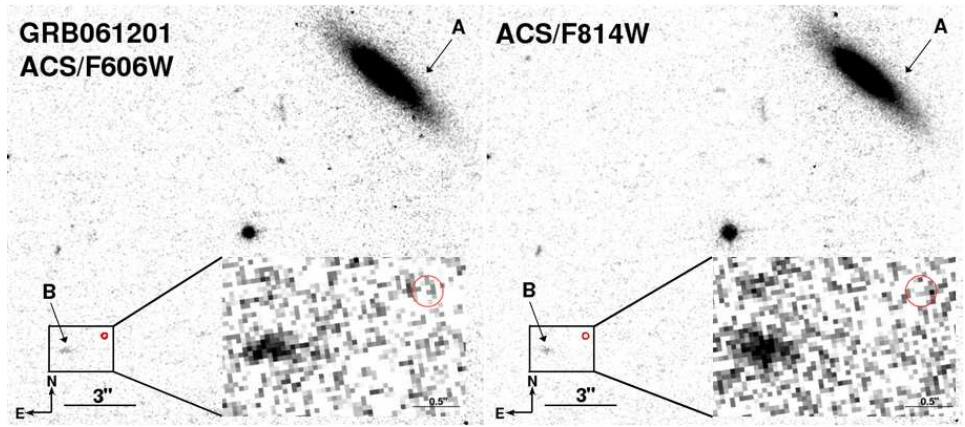}
\caption{{\it Left:} {\it HST}/ACS/F606W image of the location of
GRB\,061201.  The bright galaxy at the upper right-hand corner (``A'')
is located at $z=0.111$ \citep{gcn5952,sdp+07} with an offset of about
32.5 kpc.  A second, fainter galaxy (``B'') is located about $1.8''$
away from the optical afterglow position (see
Appendix~\ref{sec:app2}).  {\it Right:} Same, but for the {\it
HST}/ACS/F814W observations.
\label{fig:061201}}
\end{figure}

The {\it HST} observations of GRB 061201 and its environment are shown
in Figure~\ref{fig:061201}.  We explore two possibilities for the host
galaxy.  First, the burst is located $16.2''$ (32.5 kpc) from a
relatively bright galaxy at $z=0.111$ (marked ``A'' in
Figure~\ref{fig:061201}; \citealt{gcn5952,sdp+07}), for which we
measure $m({\rm F606W})=18.17$ and $m({\rm F814W})=17.82$ AB mag.
Second, we identify from the {\it HST}/ACS observations a second,
fainter galaxy (marked ``B'' in Figure~\ref{fig:061201}) located
$1.8''$ from the GRB position, and with $m({\rm F606W})=25.34$ and
$m({\rm F814W})=25.03$ AB mag.  The redshift of this galaxy is not
known, but assuming $z\gtrsim 1$ the inferred projected offset is 14.2
kpc.  The probability of chance coincidence for both galaxies is about
$20\%$ \citep{bsk+06}.  We therefore do not claim a unique host galaxy
association for this burst, and stress that both galaxies should be
considered as potential hosts.  Deeper {\it HST} observations may also
uncover an underlying host.

\section{Possible Host Galaxies of GRB\,060502b}
\label{sec:app3}

\begin{figure}
\centering
\includegraphics[angle=0,width=5.0in]{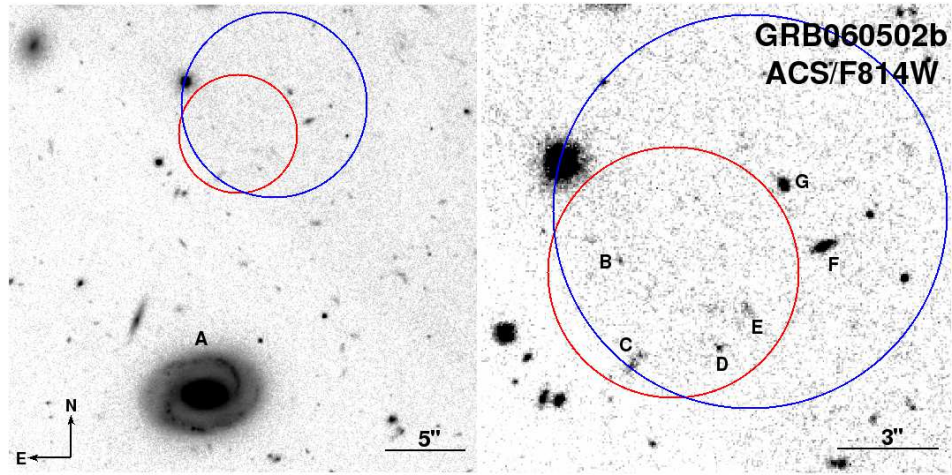}
\caption{{\it HST}/ACS/F814W images of the host galaxy of 
GRB\,060502b.  The circles mark the X-ray positions of the afterglow
from the analysis of \citet{but07} (red) and \citet{ebp+09} (blue).
The bright galaxy marked ``A'' is located at $z=0.287$ \citep{bpc+07}.
Several fainter galaxies (``B''--``G'') are located within the XRT
error circles (see Appendix~\ref{sec:app3}).
\label{fig:060502b}}
\end{figure}

The {\it HST} observations of GRB\,060502b and its environment are
shown in Figure~\ref{fig:060502b}.  Previously, \citet{bpc+07} claimed
that the host is likely an early type galaxy at $z=0.287$ located
about $70$ kpc away from the burst XRT position.  These authors also
note the presence of fainter objects within the XRT error circle.  A
galaxy with $R\approx 25.2$ mag was also found by \citet{bfp+07}.  In
the combined {\it HST}/ACS/F814W we find 6 faint galaxies within the
XRT error circles of GRB\,060502b (Figure~\ref{fig:060502b}).  These
galaxies have the following AB magnitudes: $27.5$ (B), $25.9$ (C), 
$27.2$ (D), $26.1$ (E), $24.8$ (F), $25.7$ (G).  The probability of
chance coincidence for these galaxies within the XRT error circles
is of the order of unity.

%\bibliographystyle{apj}
%\bibliography{journals_apj,refs,refs2,refs3,refs4}

\begin{thebibliography}{}

\bibitem[\protect\citeauthoryear{{Barbier} et~al.}{{Barbier}
  et~al.}{2007}]{gcn6623}
{Barbier}, L., et~al. 2007, GRB Coordinates Network, 6623, 1

\bibitem[\protect\citeauthoryear{{Barnes} \& {Hernquist}}{{Barnes} \&
  {Hernquist}}{1992}]{bh92}
{Barnes}, J.~E.,  \& {Hernquist}, L. 1992, \araa, 30, 705

\bibitem[\protect\citeauthoryear{{Beckwith} et~al.}{{Beckwith}
  et~al.}{2006}]{bsk+06}
{Beckwith}, S.~V.~W., et~al. 2006, \aj, 132, 1729

\bibitem[\protect\citeauthoryear{{Belczynski} et~al.}{{Belczynski}
  et~al.}{2006}]{bpb+06}
{Belczynski}, K., {Perna}, R., {Bulik}, T., {Kalogera}, V., {Ivanova}, N.,  \&
  {Lamb}, D.~Q. 2006, \apj, 648, 1110

\bibitem[\protect\citeauthoryear{{Berger}}{{Berger}}{2006}]{gcn5952}
{Berger}, E. 2006, GRB Coordinates Network, 5952, 1

\bibitem[\protect\citeauthoryear{{Berger}}{{Berger}}{2009}]{ber09}
{Berger}, E. 2009, \apj, 690, 231

\bibitem[\protect\citeauthoryear{{Berger} et~al.}{{Berger}
  et~al.}{2009}]{bcf+09}
{Berger}, E., {Cenko}, S.~B., {Fox}, D.~B.,  \& {Cucchiara}, A. 2009, ArXiv
  e-prints

\bibitem[\protect\citeauthoryear{{Berger} et~al.}{{Berger}
  et~al.}{2007}]{bfp+07}
{Berger}, E., et~al. 2007, \apj, 664, 1000

\bibitem[\protect\citeauthoryear{{Berger} et~al.}{{Berger}
  et~al.}{2005}]{bpc+05}
{Berger}, E., et~al. 2005, \nat, 438, 988

\bibitem[\protect\citeauthoryear{{Bloom} et~al.}{{Bloom} et~al.}{1998}]{bdk+98}
{Bloom}, J.~S., {Djorgovski}, S.~G., {Kulkarni}, S.~R.,  \& {Frail}, D.~A.
  1998, \apjl, 507, L25

\bibitem[\protect\citeauthoryear{{Bloom}, {Kulkarni}, \& {Djorgovski}}{{Bloom}
  et~al.}{2002}]{bkd02}
{Bloom}, J.~S., {Kulkarni}, S.~R.,  \& {Djorgovski}, S.~G. 2002, \aj, 123, 1111

\bibitem[\protect\citeauthoryear{{Bloom} et~al.}{{Bloom} et~al.}{2007}]{bpc+07}
{Bloom}, J.~S., et~al. 2007, \apj, 654, 878

\bibitem[\protect\citeauthoryear{{Bloom} \& {Prochaska}}{{Bloom} \&
  {Prochaska}}{2006}]{bp06}
{Bloom}, J.~S.,  \& {Prochaska}, J.~X. 2006, in American Institute of Physics
  Conference Series, Vol. 836, Gamma-Ray Bursts in the Swift Era, ed. S.~S.
  {Holt}, N.~{Gehrels}, \& J.~A. {Nousek}, 473

\bibitem[\protect\citeauthoryear{{Bloom} et~al.}{{Bloom} et~al.}{2006}]{bpp+06}
{Bloom}, J.~S., et~al. 2006, \apj, 638, 354

\bibitem[\protect\citeauthoryear{{Bloom}, {Sigurdsson}, \& {Pols}}{{Bloom}
  et~al.}{1999}]{bsp99}
{Bloom}, J.~S., {Sigurdsson}, S.,  \& {Pols}, O.~R. 1999, \mnras, 305, 763

\bibitem[\protect\citeauthoryear{{Butler}}{{Butler}}{2007}]{but07}
{Butler}, N.~R. 2007, \aj, 133, 1027

\bibitem[\protect\citeauthoryear{{Ciotti} \& {Bertin}}{{Ciotti} \&
  {Bertin}}{1999}]{cb99}
{Ciotti}, L.,  \& {Bertin}, G. 1999, \aap, 352, 447

\bibitem[\protect\citeauthoryear{{Conselice} et~al.}{{Conselice}
  et~al.}{2005}]{cvf+05}
{Conselice}, C.~J., et~al. 2005, \apj, 633, 29

\bibitem[\protect\citeauthoryear{{D'Avanzo} et~al.}{{D'Avanzo}
  et~al.}{2009}]{dmc+09}
{D'Avanzo}, P., et~al. 2009, \aap, 498, 711

\bibitem[\protect\citeauthoryear{{Djorgovski} et~al.}{{Djorgovski}
  et~al.}{1998}]{dkb+98}
{Djorgovski}, S.~G., {Kulkarni}, S.~R., {Bloom}, J.~S., {Goodrich}, R.,
  {Frail}, D.~A., {Piro}, L.,  \& {Palazzi}, E. 1998, \apjl, 508, L17

\bibitem[\protect\citeauthoryear{{Donaghy} et~al.}{{Donaghy}
  et~al.}{2006}]{dls+06}
{Donaghy}, T.~Q., et~al. 2006, ArXiv Astrophysics e-prints

\bibitem[\protect\citeauthoryear{{Eichler} et~al.}{{Eichler}
  et~al.}{1989}]{elp+89}
{Eichler}, D., {Livio}, M., {Piran}, T.,  \& {Schramm}, D.~N. 1989, \nat, 340,
  126

\bibitem[\protect\citeauthoryear{{Evans} et~al.}{{Evans} et~al.}{2009}]{ebp+09}
{Evans}, P.~A., et~al. 2009, \mnras, 397, 1177

\bibitem[\protect\citeauthoryear{{Fox} et~al.}{{Fox} et~al.}{2005}]{ffp+05}
{Fox}, D.~B., et~al. 2005, \nat, 437, 845

\bibitem[\protect\citeauthoryear{{Freeman}}{{Freeman}}{1970}]{fre70}
{Freeman}, K.~C. 1970, \apj, 160, 811

\bibitem[\protect\citeauthoryear{{Fruchter} \& {Hook}}{{Fruchter} \&
  {Hook}}{2002}]{fh02}
{Fruchter}, A.~S.,  \& {Hook}, R.~N. 2002, \pasp, 114, 144

\bibitem[\protect\citeauthoryear{{Fruchter} et~al.}{{Fruchter}
  et~al.}{2006}]{fls+06}
{Fruchter}, A.~S., et~al. 2006, \nat, 441, 463

\bibitem[\protect\citeauthoryear{{Fruchter} et~al.}{{Fruchter}
  et~al.}{1999}]{ftm+99}
{Fruchter}, A.~S., et~al. 1999, \apjl, 519, L13

\bibitem[\protect\citeauthoryear{{Fryer}, {Woosley}, \& {Hartmann}}{{Fryer}
  et~al.}{1999}]{fwh99}
{Fryer}, C.~L., {Woosley}, S.~E.,  \& {Hartmann}, D.~H. 1999, \apj, 526, 152

\bibitem[\protect\citeauthoryear{{Gal-Yam} et~al.}{{Gal-Yam}
et~al.}{2007}]{gfp+06} 
{Gal-Yam}, A., et~al. 2006, \nat, 444, 1053

\bibitem[\protect\citeauthoryear{{Gehrels} et~al.}{{Gehrels}
  et~al.}{2005}]{gso+05}
{Gehrels}, N., et~al. 2005, \nat, 437, 851

\bibitem[\protect\citeauthoryear{{Graham} et~al.}{{Graham}
  et~al.}{2009}]{gfl+09}
{Graham}, J.~F., et~al. 2009, \apj, 698, 1620

\bibitem[\protect\citeauthoryear{{Hjorth} et~al.}{{Hjorth}
  et~al.}{2005}]{hwf+05}
{Hjorth}, J., et~al. 2005, \nat, 437, 859

%\bibitem[\protect\citeauthoryear{{Jaunsen} et~al.}{{Jaunsen}
%  et~al.}{2003}]{jao+03}
%{Jaunsen}, A.~O., et~al. 2003, \aap, 402, 125

\bibitem[\protect\citeauthoryear{{Kelly}, {Kirshner}, \& {Pahre}}{{Kelly}
  et~al.}{2008}]{kkp08}
{Kelly}, P.~L., {Kirshner}, R.~P.,  \& {Pahre}, M. 2008, \apj, 687, 1201

\bibitem[\protect\citeauthoryear{{Koekemoer} et~al.}{{Koekemoer}
  et~al.}{2002}]{kfh+02}
{Koekemoer}, A.~M., {Fruchter}, A.~S., {Hook}, R.~N.,  \& {Hack}, W. 2002, in
  The 2002 HST Calibration Workshop : Hubble after the Installation of the ACS
  and the NICMOS Cooling System, ed. S.~{Arribas}, A.~{Koekemoer}, \&
  B.~{Whitmore}, 337

\bibitem[\protect\citeauthoryear{{Levan} et~al.}{{Levan}
  et~al.}{2006a}]{ltf+06}
{Levan}, A.~J., et~al. 2006a, \apjl, 648, L9

\bibitem[\protect\citeauthoryear{{Levan} et~al.}{{Levan}
  et~al.}{2006b}]{lwc+06}
{Levan}, A.~J., {Wynn}, G.~A., {Chapman}, R., {Davies}, M.~B., {King}, A.~R.,
  {Priddey}, R.~S.,  \& {Tanvir}, N.~R. 2006b, \mnras, 368, L1

\bibitem[\protect\citeauthoryear{{Malesani} et~al.}{{Malesani}
  et~al.}{2007}]{mcd+07}
{Malesani}, D., et~al. 2007, \aap, 473, 77

\bibitem[\protect\citeauthoryear{{Metzger}, {Quataert}, \&
  {Thompson}}{{Metzger} et~al.}{2008}]{mqt08}
{Metzger}, B.~D., {Quataert}, E.,  \& {Thompson}, T.~A. 2008, \mnras, 385, 1455

\bibitem[\protect\citeauthoryear{{Narayan}, {Paczynski}, \& {Piran}}{{Narayan}
  et~al.}{1992}]{npp92}
{Narayan}, R., {Paczynski}, B.,  \& {Piran}, T. 1992, \apjl, 395, L83

\bibitem[\protect\citeauthoryear{{Ofek} et~al.}{{Ofek}
et~al.}{2007}]{ocg+07} 
{Ofek}, E.~O., et~al. 2007, \apj, 662, 1129

\bibitem[\protect\citeauthoryear{{Peng} et~al.}{{Peng} et~al.}{2007}]{phi+07}
{Peng}, C.~Y., {Ho}, L.~C., {Impey}, C.~D.,  \& {Rix}, H.~W. 2007, in Bulletin
  of the American Astronomical Society, Vol.~38, Bulletin of the American
  Astronomical Society, 804

\bibitem[\protect\citeauthoryear{{Perley} et~al.}{{Perley}
  et~al.}{2008}]{pbm+08}
{Perley}, D.~A., {Bloom}, J.~S., {Modjaz}, M., {Miller}, A.~A., {Shiode}, J.,
  {Brewer}, J., {Starr}, D.,  \& {Kennedy}, R. 2008, GRB Coordinates Network,
  7889, 1

\bibitem[\protect\citeauthoryear{{Perley} et~al.}{{Perley}
  et~al.}{2009}]{pmg+09}
{Perley}, D.~A., et~al. 2009, \apj, 696, 1871

\bibitem[\protect\citeauthoryear{{Piranomonte} et~al.}{{Piranomonte}
  et~al.}{2008}]{pdc+08}
{Piranomonte}, S., et~al. 2008, \aap, 491, 183

\bibitem[\protect\citeauthoryear{{Rau}, {McBreen}, \& {Kruehler}}{{Rau}
  et~al.}{2009}]{gcn9353}
{Rau}, A., {McBreen}, S.,  \& {Kruehler}, T. 2009, GRB Coordinates Network,
  9353, 1

\bibitem[\protect\citeauthoryear{{Sakamoto} et~al.}{{Sakamoto}
  et~al.}{2007}]{gcn7156}
{Sakamoto}, T., {Norris}, J., {Ukwatta}, T., {Barthelmy}, S.~D., {Gehrels}, N.,
   \& {Stamatikos}, M. 2007, GRB Coordinates Network, 7156, 1

\bibitem[\protect\citeauthoryear{{Schlegel}, {Finkbeiner}, \&
 {Davis}}{{Schlegel} et~al.}{1998}]{sfd98} {Schlegel}, D.~J., {Finkbeiner},
 D.~P., \& {Davis}, M. 1998, \apj, 500, 525

\bibitem[\protect\citeauthoryear{{Soderberg} et~al.}{{Soderberg}
  et~al.}{2006}]{sbk+06}
{Soderberg}, A.~M., et~al. 2006, \apj, 650, 261

\bibitem[\protect\citeauthoryear{{Stratta} et~al.}{{Stratta}
  et~al.}{2007}]{sdp+07}
{Stratta}, G., et~al. 2007, \aap, 474, 827

\bibitem[\protect\citeauthoryear{{Troja} et~al.}{{Troja} et~al.}{2008}]{tko+08}
{Troja}, E., {King}, A.~R., {O'Brien}, P.~T., {Lyons}, N.,  \& {Cusumano}, G.
  2008, \mnras, 385, L10

\bibitem[\protect\citeauthoryear{{Trujillo} et~al.}{{Trujillo}
  et~al.}{2004}]{tea+04}
{Trujillo}, I., {Erwin}, P., {Asensio Ramos}, A.,  \& {Graham}, A.~W. 2004,
  \aj, 127, 1917

\bibitem[\protect\citeauthoryear{{Ukwatta} et~al.}{{Ukwatta}
  et~al.}{2009}]{gcn9337}
{Ukwatta}, T.~N., et~al. 2009, GRB Coordinates Network, 9337, 1

\bibitem[\protect\citeauthoryear{{van den Bergh}, {Li}, \&
{Filippenko}}{{van den Bergh} et~al.}{2005}]{vlf05} {van den Bergh},
S., {Li}, W., \& {Filippenko}, A.~V.  2005, PASP, 117, 773

\bibitem[\protect\citeauthoryear{{Wainwright}, {Berger}, \&
  {Penprase}}{{Wainwright} et~al.}{2007}]{wbp07}
{Wainwright}, C., {Berger}, E.,  \& {Penprase}, B.~E. 2007, \apj, 657, 367

\end{thebibliography}

\clearpage
\begin{deluxetable}{lccccccccccc}
%\rotate
\tabletypesize{\scriptsize}
\tablecolumns{12}
\tabcolsep0.0in\footnotesize
\tablewidth{0pc}
\tablecaption{{\it HST} Observations of Short GRB Host Galaxies
\label{tab:hst}}
\tablehead{
\colhead{GRB}                  &
\colhead{RA}                   &
\colhead{Dec}                  &
\colhead{Uncert.}              &
\colhead{OA?}                  &
\colhead{$z$}                  &
\colhead{Instrument}           &
\colhead{Filter}               &
\colhead{Date}                 &
\colhead{Exp.~Time}            &
\colhead{AB mag$\,^a$}         &
\colhead{$A_{\lambda}\,^b$}    \\           
\colhead{}                     &
\colhead{(J2000)}              &
\colhead{(J2000)}              &
\colhead{($''$)}               &
\colhead{}                     &
\colhead{}                     &
\colhead{}                     &
\colhead{}                     &
\colhead{(UT)}                 &
\colhead{(s)}                  &
\colhead{}                     &
\colhead{(mag)}                           
}
\startdata
050509b & \ra{12}{36}{14.06} & \dec{+28}{59}{07.2}$\,^c$ & 3.4 & N & 0.226 & ACS    & F814W & 2005 May 14 & 6870 & 16.32 & 0.037  \\
        & \ra{12}{36}{13.76} & \dec{+28}{59}{03.2}       & 3.3 &   &       &        &       &             &      &       &        \\
\\
050709  & \ra{23}{01}{26.96} & \dec{-38}{58}{39.5} & 0.2 & Y & 0.1606 & ACS   & F814W & 2006 Jul 16 & 6981 & 21.09 & 0.02 \\
        &                    &                     &     &   &        & WFPC2 & F450W & 2007 Jul 29 & 3200 & 21.43 & 0.045 \\
\\
050724  & \ra{16}{24}{44.38} & \dec{-27}{32}{27.5} & 0.1 & Y & 0.257 & WFPC2 & F450W & 2008 Apr 07 & 3200 & 19.98 & 2.645 \\
        &                    &                     &     &   &       & WFPC2 & F814W & 2008 May 18 & 3200 & 18.74 & 1.189 \\
\\
051210  & \ra{22}{00}{41.26} & \dec{-57}{36}{46.5} & 2.9 & N & $>1.4$ & WFPC2 & F675W & 2007 Apr 03 & 2800 & 21.14 & 0.052 \\
        & \ra{22}{00}{41.33} & \dec{-57}{36}{49.4} & 1.7 &   &        &       &       &             &      &       &       \\
\\
051221a & \ra{21}{54}{48.63} & \dec{+16}{53}{27.4} & 0.2 & Y & 0.5465 & WFPC2 & F555W & 2007 Aug 13 & 3200 & 21.86 & 0.227 \\
        &                    &                     &     &   &        & WFPC2 & F814W & 2007 Aug 22 & 1600 & 21.42 & 0.133 \\
\\                                                                             
060121  & \ra{09}{09}{51.99} & \dec{+45}{39}{45.6} & 0.1 & Y & \nod & ACS & F606W & 2006 Feb 27 & 4400 & 26.22 & 0.047 \\
\\
060313  & \ra{04}{26}{28.42} & \dec{-10}{50}{39.9} & 0.2 & Y & \nod & ACS & F475W & 2006 Oct 13 & 2088 & 26.38 & 0.300 \\
        &                    &                     &     &   &      & ACS & F775W & 2006 Oct 14 & 2120 & 25.61 & 0.135 \\
\\
060502b & \ra{18}{35}{45.53} & \dec{+52}{37}{52.9} & 3.7 & N & \nod & ACS & F814W & 2006 May 15-Jul 16 & 25224 & 17.88 / 24.8-27.5$\,^d$ & 0.085 \\
        & \ra{18}{35}{45.28} & \dec{+52}{37}{54.7} & 5.8 &   &      &     &       &                    &       &                  &       \\
\\
061006  & \ra{07}{24}{07.78} & \dec{-79}{11}{55.5} & 0.2 & Y & 0.4377 & ACS   & F814W & 2006 Oct 14 & 6054 & 21.67 & 0.616 \\
        &                    &                     &     &   &        & WFPC2 & F555W & 2008 May 22 & 3200 & 23.90 & 1.052 \\
\\
061201  & \ra{22}{08}{32.09} & \dec{-74}{34}{47.1} & 0.2 & Y & 0.111 / \nod$\,^e$ & ACS & F606W & 2006 Dec 11 & 2178 & 18.17 / 25.34$\,^f$ & 0.251 \\
        &                    &                     &     &   &                    & ACS & F814W & 2006 Dec 11 & 2244 & 17.82 / 25.03$\,^f$ & 0.147
\enddata
\tablecomments{Summary of short GRB positions and redshifts, {\it HST}
observations, and host galaxy magnitudes (calculated using IRAF/{\tt
ellipse}). \\
$^a$ These values have been corrected for Galactic extinction. \\
$^b$ Galactic extinction. \\
$^c$ In all cases with {\it Swift}/XRT positions the top and bottom 
set of coordinates are from the catalogs of \citet{but07} and 
\citet{ebp+09}, respectively. \\
$^d$ These magnitudes correspond to galaxy ``A'' and galaxies ``B''--
``G'' in Figure~\ref{fig:060502b}. \\
$^e$ We consider two possible host galaxies for this burst (see
Appendix~\ref{sec:app2}). \\
$^f$ The first value is for galaxy ``A'' and the second is for galaxy
``B'' in Figure~\ref{fig:061201}.
}
\end{deluxetable}

\clearpage
\begin{deluxetable}{lcccccccccccc}
\tabletypesize{\scriptsize}
\tablecolumns{13}
\tabcolsep0.0in\footnotesize
\tablewidth{0pc}
\tablecaption{Morphological Properties of Short GRB Host Galaxies
\label{tab:morph}}
\tablehead{
\colhead{}			 &
\colhead{}			 &
\colhead{}			 &
\multicolumn{5}{c}{{\tt galfit}} &
\colhead{}			 &
\multicolumn{4}{c}{IRAF/{\tt ellipse}} \\
\cline{4-8}\cline{10-13} 	\\
\colhead{GRB}			&
\colhead{Instrument}		&
\colhead{Filter}		&
\colhead{$n\,^a$}		&
\colhead{$r_e$}		        &
\colhead{$r_e\,^c$}		&
\colhead{$\mu_e\,^b$}		&
\colhead{AB Mag$\,^b$}          &
\colhead{}			&
\colhead{$n$}			&
\colhead{$r_e$}		        &
\colhead{$r_e\,^c$}		&
\colhead{$\mu_e\,^b$}		\\
\colhead{}			&
\colhead{}			&
\colhead{}			&
\colhead{}			&
\colhead{($''$)}		&
\colhead{(kpc)}		        &
\colhead{(AB mag arcsec$^{-2}$)} &
\colhead{}			&
\colhead{}			&
\colhead{}			&
\colhead{($''$)}		&
\colhead{(kpc)}		        &
\colhead{(AB mag arcsec$^{-2}$)}		
}
\startdata
050509b & ACS & F814W   & 5.6 & 5.84 & 20.98 & 23.5 & 16.2 & & 5.6 & 5.84 & 20.98 & 23.4 \\
\\
050709  & ACS   & F814W & 1.1 & 0.76 & 2.08 & 23.0 & 21.2 & & 0.6 & 0.64 & 1.75 & 22.4 \\
	& WFPC2 & F450W & 1.1 & 0.71 & 1.94 & 23.5 & 21.9 & & 0.9 & 0.71 & 1.94 & 23.4 \\
\\
050724  & WFPC2 & F450W & 4 & 1.35 & 5.34 & 23.2 & 19.4 & & 1.3 & 0.36 & 1.42 & 20.8 \\
	& WFPC2 & F814W & 3.0 & 0.82 & 3.24 & 20.5 & 18.0 & & 2.9 & 1.01 & 4.00 & 20.8 \\
\\
051210  & WFPC2 & F675W & 1 & 0.70 & 5.63 & 23.8 & 23.7 & & 1.0 & 0.63 & 5.07 & 24.2  \\
	&       &       & 4 & 2.38 & 19.14 & 26.3 & 22.8 \\
\\
051221a & WFPC2 & F555W & 0.9 & 0.36 & 2.29 & 23.3 & 23.1 & & 0.8 & 0.34 & 2.17 & 23.1 \\
	& WFPC2 & F814W & 0.9 & 0.41 & 2.61 & 22.7 & 22.1 & & 0.9 & 0.39 & 2.49 & 22.7 \\
\\	     	 		    
060121  & ACS & F606W   & 1 & 0.36 & 2.89 & 25.9 & 27.1 & & 1.4 & 0.67 & 5.39 & 27.2 \\
	&     &         & 4 & 1.22 & 9.81 & 27.4 & 26.6 \\
\\
060313  & ACS & F475W   & 1 & 0.14 & 1.13 & 23.7 & 27.3 & & 0.6 & 0.17 & 1.37 & 24.9 \\
	&     &         & 4 & 0.32 & 2.57 & 26.2 & 26.7 \\
	& ACS & F775W   & 1 & 0.07 & 0.56 & 21.4 & 26.1 & & 1.3 & 0.23 & 1.85 & 25.0 \\
	&     &         & 4 & 0.10 & 0.80 & 23.6 & 26.3 \\
\\
061006  & ACS   & F814W & 0.7 & 0.57 & 3.22 & 22.3 & 22.7 & & 0.7 & 0.65 & 3.67 & 22.9 \\
	& WFPC2 & F555W & 1   & 0.63 & 3.55 & 23.3 & 23.4 & & 0.8 & 0.55 & 3.10 & 23.6
\enddata
\tablecomments{Results of morphological analysis performed with {\tt
galfit} and IRAF/{\tt ellipse} (\S\ref{sec:morph}).\\
$^a$ For the S\'{e}rsic index $n$ in {\tt galfit}, exact values of 1 
and 4 indicate a fit with $n$ as a fixed parameter. \\
$^b$ These values have been corrected for Galactic extinction. \\
$^c$ For the hosts with unknown redshift (GRBs 051210, 060121, 
and 060313) we assume $z=1$.}
\end{deluxetable}

\clearpage
\begin{deluxetable}{lccccccccccccc}
\tabletypesize{\scriptsize}
%\rotate
\tablecolumns{14}
\tabcolsep0.0in\footnotesize
\tablewidth{0pc}
\tablecaption{Short GRB Angular, Physical, and Host-Normalized Offsets
\label{tab:offsets}}
\tablehead {
\colhead {GRB}		&
\colhead {Instrument}	&
\colhead {Filter}	&
\colhead {$z$}		&
\colhead {Reference} 		&
\colhead {No.}		&
\colhead {$\sigma_{\rm GB\rightarrow HST}$} 	&
\colhead {$\sigma_{\rm \theta,GRB}$} 		&
\colhead {$\sigma_{\rm \theta,gal}$} 		&
\colhead {$\delta$RA}	&
\colhead {$\delta$Dec}	&
\colhead {Offset}	&
\colhead {Offset}	&             
\colhead {Offset}	\\
\colhead {}		&
\colhead {}		&
\colhead {}		&
\colhead {}		&
\colhead {}		&
\colhead {}		&
\colhead {(mas)}	&
\colhead {(mas)}	&
\colhead {(mas)}	&
\colhead {($''$)}	&
\colhead {($''$)}	&
\colhead {($''$)}	&
\colhead {(kpc)}	&
\colhead {($r_e\,^a$)}	
}
\startdata
050509b & ACS & F814W & 0.226 & SDSS & 10 & 30 & 3400 & 1.0 & +15.61 & +8.40 & $17.73\pm 3.4$ & $63.7\pm 12.2$ & $3.04\pm 0.58$ \\
        &     &       &       &      &    &    & 3300 & 1.0 & +11.68 & +4.40 & $12.48\pm 3.3$ & $44.8\pm 11.9$ & $2.14\pm 0.57$ \\
\\
050709 & ACS   & F814W & 0.1606 & HST/ACS & 35   & 8    & 1.0 & 1.4$\,^b$ & +1.294 & $-0.310$ & $1.331\pm 0.010$ & $3.64\pm 0.027$ & $2.08\pm 0.02$ \\
       & WFPC2 & F450W &        & HST/ACS & 12   & 14   & 1.0 & 4.7 & +1.306 & $-0.360$       & $1.355\pm 0.020$ & $3.71\pm 0.055$ & $1.91\pm 0.03$ \\
       & ACS   & F814W &        & self    & \nod & \nod & 1.0 & 1.4 & +1.329 & $-0.310$       & $1.365\pm 0.002$ & $3.74\pm 0.005$ & $2.13\pm 0.01$ \\
\\
050724 & WFPC2 & F450W   & 0.257 & Magellan/PANIC & 60   & 15   & 5.0 & 4.7$\,^c$ & $-0.226$ & $-0.640$ & $0.679\pm 0.025$ & $2.69\pm 0.099$ & $1.89\pm 0.07$ \\
       & WFPC2 & F814W   &       & Magellan/PANIC & 85   & 14   & 5.0 & 1.4       & $-0.213$ & $-0.630$ & $0.665\pm 0.020$ & $2.63\pm 0.079$ & $0.66\pm 0.02$ \\
       & PANIC & $K_{s}$ &       & self           & \nod & \nod & 5.0 & 1.0       & $-0.253$ & $-0.650$ & $0.697\pm 0.006$ & $2.76\pm 0.024$ & \nod \\
\\
051210 & WFPC2 & F675W & $>1.4$ & 2MASS & 12 & 29 & 2900 & 8 & +2.89 & +0.50   & $2.93\pm 2.9$ & $24.9\pm 24.6$ & $4.65\pm 4.60$ \\
       &       &       &        &       &    &    & 1700 & 8 & +3.45 & $-2.40$ & $4.20\pm 1.7$ & $35.7\pm 14.4$ & $6.67\pm 2.70$ \\
\\
051221 & WFPC2 & F555W & 0.5465 & Gemini-N/GMOS & 45 & 23 & 2.5 & 3.1 & $-0.287$ & +0.090 & $0.301\pm 0.029$ & $1.92\pm 0.18$ & $0.88\pm 0.08$ \\
       & WFPC2 & F814W &        & Gemini-N/GMOS & 45 & 24 & 2.5 & 3.1 & $-0.330$ & +0.090 & $0.342\pm 0.030$ & $2.18\pm 0.19$ & $0.88\pm 0.08$ \\
\\
060121 & ACS & F606W & \nod & Gemini-N/GMOS & 25 & 18 & 16 & 12 & $-0.115$ & +0.030& $0.119\pm 0.046$ & $0.96\pm 0.37\,^d$ & $0.18\pm 0.07$ \\
\\
060313 & ACS & F475W & \nod & Gemini-S/GMOS & 30 & 30 & 19 & 19.4 & +0.354 & +0.040 & $0.356\pm 0.068$ & $2.86\pm 0.55\,^d$ & $2.09\pm 0.40$ \\
       & ACS & F775W &      & Gemini-S/GMOS & 15 & 30 & 19 & 13.2 & +0.280 & +0.050 & $0.284\pm 0.062$ & $2.28\pm 0.50\,^d$ & $1.23\pm 0.23$ \\
\\
060502b & ACS & F814W &     & USNO-B        & 47 & 120 & 3700 & \nod$\,^e$ & \nod & \nod & \nod & \nod & \nod \\
        &     &       &     &               &    &     & 5800 & \nod$\,^e$ & \nod & \nod & \nod & \nod & \nod \\
\\
061006 & ACS   & F814W & 0.4377 & VLT/FORS1 & 75 & 17 & 21 & 4.9 & $-0.155$  & $-0.170$ & $0.230\pm 0.043$ & $1.30\pm 0.24$ & $0.35\pm 0.07$ \\
       & WFPC2 & F555W &        & VLT/FORS1 & 45 & 20 & 21 & 11  & $-0.171$  & $-0.190$ & $0.256\pm 0.052$ & $1.44\pm 0.29$ & $0.46\pm 0.10$ \\
\\
061201 & ACS & F814W & \nod & VLT/FORS2 & 24 & 13 & 41 & \nod$^e$ & \nod & \nod & \nod & \nod & \nod \\
       & ACS & F606W & \nod & VLT/FORS2 & 24 & 13 & 41 & \nod$^e$ & \nod & \nod & \nod & \nod & \nod
\enddata
\tablecomments{Projected angular, physical, and host-normalized offsets
for the short GRBs with {\it HST} observations.  \\
$^a$ Values for $r_e$ are from {\tt ellipse} (Table~\ref{tab:morph}). \\
$^b$ Systematic uncertainty in host center is ~50 mas. \\
$^c$ Systematic uncertainty in host center is ~20 mas. \\
$^d$ Assuming $z=1$. \\
$^e$ We do not claim a unique host galaxy identification for this burst.}
\end{deluxetable}
%\end{landscape}

\clearpage
\begin{deluxetable}{lccc}
\tabletypesize{\scriptsize}
\tablecolumns{4}
\tabcolsep0.0in\footnotesize
\tablewidth{0pc}
\tablecaption{Short GRB Fractional Flux
\label{tab:lightfrac}}
\tablehead {
\colhead {GRB}		&
\colhead {Instrument}	&
\colhead {Filter}	&
\colhead {Fractional Flux}	
}
\startdata
050709 & WFPC2 & F450W & 0    \\
       & ACS   & F814W & 0.09 \\
\\
050724 & WFPC2 & F450W & 0.03 \\
       & WFPC2 & F814W & 0.33 \\
\\
051221 & WFPC2 & F555W & 0.54 \\
       & WFPC2 & F814W & 0.65 \\
\\
060121 & ACS   & F606W & 0.41 \\
\\
060313 & ACS   & F475W & 0.04 \\
       & ACS   & F775W & 0    \\
\\
061006 & WFPC2 & F555W & 0.56 \\
       & ACS   & F814W & 0.63 \\
\enddata
\tablecomments{Fraction of host galaxy light in pixels fainter than
the GRB position.}
\end{deluxetable}

\clearpage

\clearpage

\clearpage

\clearpage

\clearpage

\clearpage

\clearpage

\clearpage

\clearpage

\clearpage

\clearpage

\clearpage

\clearpage

\clearpage

\clearpage

\clearpage

\end{document}